%% file: manuscript.tex
\documentclass[superscriptaddress,amsmath,amssymb,aps,prd,twocolumn,tightenlines,nofootinbib]{revtex4-1}

\pdfoutput=1

\usepackage{graphicx}
\usepackage{xcolor}

\usepackage{dsfont}
\usepackage{graphicx} 
\usepackage{placeins}
\usepackage{epsfig}
\usepackage{hyperref}
\usepackage{ulem}

\newcommand{\orcid}[1]{\href{https://orcid.org/#1}{\textcolor[HTML]{A6CE39}{\aiOrcid}}}

\newcommand{\be}{\begin{equation}}
\newcommand{\ee}{\end{equation}}

\def\bi{\begin{itemize}
}
\def\ei{\end{itemize}}

\begin{document}

\title{Frequentist versus Bayesian analyses:
Cross-correlation as an (approximate) 
sufficient 
statistic for LIGO-Virgo stochastic background searches}
\author{Andrew Matas}
\email{andrew.matas@aei.mpg.de}
\affiliation{Max Planck Institute for Gravitational Physics 
(Albert Einstein Institute),
D-14476 Potsdam, Germany}
\author{Joseph D.~Romano} 
\email{joseph.d.romano@ttu.edu} 
\affiliation{Department of Physics and Astronomy,
Texas Tech University, Box 41051,
Lubbock, TX 79409-1051, USA}
\date{\today}

\input{abstract.tex}

\maketitle

\input{intro.tex}

\input{example.tex}

\input{crosscorr_white.tex}
\input{crosscorr_colored.tex}

\input{discussion.tex}

%%%%%%%%%%%%%%%%%%%%%%%%%%
\begin{acknowledgments}

We thank Stephen Taylor for useful discussions about Bayesian analyses 
and noise-marginalized statistics in the context of pulsar timing arrays.
We also thank Colm Talbot and Sylvia Biscoveanu for conversations at the
early stages of this work, regarding fully Bayesian searches for 
stochastic backgrounds using LIGO-Virgo data.  
JDR thanks Tom Callister, Michael Coughlin, Joey Key, and Tyson 
Littenberg for many useful discussions, which took place several years 
ago and which eventually led to this
project, albeit in a slightly different form than originally planned.
JDR also acknowledges support of start-up funds from Texas Tech University. 
The authors are grateful for computational resources provided by the LIGO Laboratory and supported by National Science Foundation Grants PHY-0757058 and PHY-0823459.
Numerical computations were performed with the \texttt{numpy} \cite{numpy}, \texttt{scipy}
\cite{2020SciPy-NMeth}, and \texttt{bilby} \cite{Ashton:2018jfp} libraries, and plots with the \texttt{matplotlib} \cite{Hunter:2007}
library. 

\end{acknowledgments}

\appendix
\input{appA}

\input{appB}

\input{appC}

%%%%%%%%%%%%%%%%%%%%%%%%%
\bibliographystyle{apsrev4-1}
\bibliography{refs}

\end{document}

%% file: abstract.tex
\begin{abstract}
Sufficient statistics are combinations of data in terms of 
which the likelihood function can be rewritten without loss 
of information.
Depending on the data volume reduction, the 
use of sufficient statistics as a preliminary step in a 
Bayesian analysis can lead to significant increases in
efficiency when sampling from posterior distributions of
model parameters.
Here we show that the frequency integrand of the 
cross-correlation statistic and its variance are 
approximate
sufficient statistics for ground-based searches for 
stochastic gravitational-wave backgrounds.
The sufficient statistics are approximate because one
works in the weak-signal approximation and 
uses measured estimates of the auto-correlated power in
each detector.
Using analytic and numerical calculations,
we prove that LIGO-Virgo's hybrid frequentist-Bayesian
parameter estimation analysis is equivalent to a fully Bayesian
analysis. This work closes a gap in the LIGO-Virgo literature,
and suggests directions for additional searches.
\end{abstract}

%% file: intro.tex
\section{Introduction}
\label{s:intro}

Current searches for stochastic gravitational-wave backgrounds (GWBs)
using ground-based laser interferometers 
(e.g., the Advanced LIGO~\cite{aLIGO_2015} and Virgo~\cite{aVirgo_2015}
detectors) use a hybrid of frequentist and Bayesian analysis 
techniques~\cite{stoch_O2, stoch_dir_O2, RomanoCornish}.
Certain frequentist statistics 
(namely, frequency integrands for the cross-correlation statistic and its 
variance) are calculated for relatively short stretches of time-series
data (of order a couple of minutes), and then combined using inverse-noise 
weighting to produce two final frequency series, valid for the 
whole observation period (of order several months to a year).
These frequency series are then used as the fundamental input data and 
variance for a subsequent Bayesian
analysis that calculates posterior probability distributions and 
Bayesian upper limits on the strength of potential correlated 
gravitational-wave signals~\cite{StochPE}.
These hybrid frequentist-Bayesian analyses have been used 
to place upper limits on
GWBs with different amplitudes and spectral shapes \cite{stoch_O1,stoch_O2};
GWBs having non-GR polarizations predicted by alternative 
theories of gravity~\cite{TestingGR_stoch, stoch_O2}; and
potential contamination from correlated global magnetic 
fields---i.e., Schumann resonances~\cite{Schumann_1, Schumann_2, Schumann_3, Schumann_4, Meyers:2020qrb}.

On the other hand, most searches for GWBs using pulsar timing 
data~\cite{Arzoumanian:2018saf, arzoumanian2020nanograv}
and proposed searches using space-based detectors like 
LISA~\cite{Adams:2013qma} 
are {\it fully} Bayesian, proceeding directly from the time-series 
data to posterior distributions and upper limits, without ever 
calculating a cross-correlation frequency integrand or statistic.
So the question naturally arises as to whether LIGO and Virgo's 
current hybrid frequentist-Bayesian analysis is losing 
information relative to a fully-Bayesian search.

As we shall show below, the answer is basically ``no".
We assume that we can work in the weak-signal approximation, 
and we use measured 
estimates of the auto-correlated power in each detector,
as opposed to trying to infer the noise power spectra as
part of the full analysis.
Under these simplifications, the frequency integrands of the cross-correlation statistic 
and its variance are approximately lossless combinations of 
the full time-series data in terms of which the full 
likelihood function can be rewritten.
Hence, Bayesian posterior distributions produced from these
frequency series 
agree quite well with those produced from the full time-series data.
Said another way, the frequency integrands of the cross-correlation
statistic and its variance are approximate {\it sufficient 
statistics} for the analysis, so the hybrid frequentist-Bayesian 
method is essentially equivalent to a full Bayesian analysis.

The calculations presented in this paper can also be thought of as 
a providing an alternative conceptual starting point for 
LIGO-Virgo's stochastic cross-correlation analysis.
We use the full Bayesian likelihood function
as a {\it organizing} priniciple, out of which the LIGO-Virgo
stochastic analyses follows.
Various steps in the stochastic analysis, 
such as coarse graining of cross-correlated data~\cite{stoch_S1}, 
estimating auto-correlated power spectra from neighboring data 
segments~\cite{Lazzarini:bias}, 
inverse-noise weighting, optimal filtering~\cite{Allen_Romano_1999}, 
the inclusion of ``bias" factors for the 
theoretical variance~\cite{Lazzarini:bias}, 
and the use of
the cross-correlation frequency integrand and its variance to do
parameter estimation and model selection~\cite{StochPE}, all follow directly 
from the Bayesian likelihood under the assumptions of a weak signal 
and estimated auto-correlated detector power spectra.
We also see opportunities to develop new methods to look for 
signals which violate our assumptions.

Our main result will be to show that the full Gaussian likelihood
for a stochastic background 
(cf.~equations \eqref{e:product-likelihood-colored-segment} 
and \eqref{e:likelihood-colored-segment})
is equivalent to the reduced likelihood given in 
equation \eqref{e:likefinal_colored_reduced}, 
under the approximations enumerated above. 
To do this, we will build up the tools necessary in a series of increasingly more realistic (and more complex) scenarios.
In Sec.~\ref{s:example}, we give a simple example of a sufficient
statistic in the context of a search for a constant deterministic 
signal in the output of a single detector.
This example serves as a basis 
for the calculations in the following two sections
for cross-correlation-based searches for stochastic gravitational waves.
We first restrict attention in Sec.~\ref{s:CCsearches_white} to 
white signal+noise models, and stationary data.
We then extend our analyses in Sec.~\ref{s:CCsearches_colored} to
colored signal+noise models, allowing for non-stationary noise.
Both of these sections present results of different analyses 
performed on simulated data comparing posterior distributions 
produced by fully-Bayesian and sufficient-statistic analyses.
We conclude in Sec.~\ref{s:discussion} with a brief summary, 
followed by a discussion of other related approaches in the 
literature, and possible extensions of these results. 
We also include three appendices:
Appendix~\ref{s:useful_identity} contains a simple, yet very useful,
identity \eqref{e:xopt_identity} that we use repeatedly throughout 
the paper;
Appendix~\ref{s:psd_uncertainties} summarizes uncertainties in power
spectrum estimation; and 
Appendix~\ref{s:alternative_derivation} gives an alternative 
derivation of a reduced
likelihood function, but which makes different assumptions
than those given in the main text.

%% file: example.tex
\section{Simple example}
\label{s:example}

Perhaps the simplest example of a non-trivial sufficient 
statistic is the sample mean of the data 
for a constant signal in white, Gaussian noise 
(see Sec.~3.5 of Ref.~\cite{RomanoCornish} for 
a more detailed treatment of this example.)
We suppose we record $N$ time-series data samples
$d\equiv\{d_i\}$ as
\be
d_i = a + n_i\,,
\qquad i=1,2,\cdots, N\,,
\ee
where $a>0$ is the amplitude of the signal and $n_i$ 
denotes the $i$th sample of the noise.
For simplicity, we will assume that the noise is white and has 
zero mean---i.e., 
$\langle n_i \rangle =0$, $\langle n_i n_j \rangle = \sigma^2\,\delta_{ij}$,
and that the variance $\sigma^2$ is known a~priori.
The likelihood function is then
\be
p(d|a) = \frac{1}{(2\pi)^{N/2} \sigma^N}
\exp\left[-\frac{1}{2\sigma^2}\sum_i (d_i - a)^2\right]\,,
\label{e:p(d|a)}
\ee
which has the interpretation of being the probability of 
observing the data $d$ given a signal of amplitude $a$ in 
white noise with known variance $\sigma^2$.

It is fairly straight-forward to show that the 
maximum-likelihood estimator of the amplitude $a$ is 
given by the sample mean of the data
\be
\hat a
\equiv a_{\rm ML}(d) 
= \frac{1}{N}\sum_i d_i\,.
%\equiv \bar d\,.
\label{e:ahat}
\ee
This is an unbiased estimator of $a$ 
and has variance 
\be
\sigma_{\hat a}^2 =\sigma^2/N\,.
\ee
In terms of $\hat a$, 
the data-dependent term in the likelihood becomes
\be
\sum_i (d_i - a)^2 
=\sum_i d_i^2 - N\hat a^2 + N(\hat a -a)^2\,.
\ee
This equation is a special case of the general identity \eqref{e:xopt_identity}, which 
is discussed and proven in Appendix~\ref{s:useful_identity}, 
and which we will use in future sections.
The likelihood can then be rewritten in the form
\begin{multline}
p(d|a) = \frac{1}{(2\pi)^{N/2} \sigma^N}
\\
\times\exp\left[-\frac{1}{2\sigma^2}\sum_i d_i^2\right]
\exp\left[\frac{\hat{a}^2}{2\sigma_{\hat a}^2}\right]
\exp\left[-\frac{(\hat{a} - a)^2}{2\sigma^2_{\hat a}}\right]
\label{e:p(d|a)_suffstat}
\end{multline}
In this expression, we see that the data appear {\it only} 
via the combination $\hat{a}$, up to a proportionality factor 
which is {\it independent} of the parameter $a$. 
By precomputing the sample mean $\hat{a}$ and $\sum_i d_i^2$,
we can reduce the evaluation of the likelihood from $O(N)$ 
to $O(1)$ operations.

The posterior distribution for $a$, denoted $p(a|d)$, is
calculated using Bayes' theorem
\be
p(a|d) = \frac{p(d|a)p(a)}{p(d)}\,,
\ee
where $p(a)$ is the {\it prior probability distribution} 
for $a$, and 
\be
p(d) \equiv \int {\rm d}a\> p(d|a) p(a)
\ee
is the so-called {\it evidence} (or {\it marginalized likelihood}).
Since $p(d)$ is independent of $a$, the evidence acts as a normalization factor 
as far as the posterior distribution of 
$a$ is concerned.  
Thus, one often writes $p(a|d)\propto p(d|a) p(a)$ for 
posterior distribution calculations.
For this example, 
\be
p(a|d) 
\propto
\exp\left[-\frac{(\hat a-a)^2}{2\sigma_{\hat a}^2}\right]\, p(a)\,,
\ee
where the data enter the RHS of the above expression only
via the expression \eqref{e:ahat} for the maximum-likelihood 
estimator $\hat a\equiv a_{\rm ML}(d)$.
This shows that $\hat a$ is a sufficient statistic for $a$.

Finally, we note that the prior probability distribution $p(a)$ 
is chosen based on expectations for `$a$' prior to observing the data. 
It is common to use {\it flat} or {\it log-uniform} priors,
\be
p(a) = {\rm const}
\quad{\rm or}\quad
p(a) = {\rm const}/a\,,
\ee
defined over some interval
$[a_{\rm min}, a_{\rm max}]$, where the constants in 
these expressions are determined 
by the normalization condition $\int {\rm d}a\> p(a)=1$.
Note that non-trivial priors imply that the maximum-likelihood 
and maximum-posterior values will differ in general.
For the analyses that we will perform in the following sections, 
we will consider flat priors for simplicity, but 
note here that the results we obtain are valid for arbitrary,
non-flat priors as well.

%% file: crosscorr_white.tex
\section{Sufficient statistics for cross-correlation searches 
-- white signal+noise models}
\label{s:CCsearches_white}

Here we extend the calculations of the previous section to 
cross-correlation searches for stochastic gravitational waves.
We restrict our attention in Sec.~\ref{s:white} to a 
white signal+noise model, assuming stationary signal and noise
(see also Sec.~4 of Ref.~\cite{RomanoCornish}).
In Sec.~\ref{s:white_reduced}, we consider a {\it reduced} 
version of this model, where we assume weak signals relative 
to the detector noise and use measured estimates of the 
auto-correlated power in each detector, as opposed to 
having these as model parameters to be determined by the search.
Our analysis of the white noise case contains all of the
important steps present in the colored noise case that we 
shall discuss in Sec.~\ref{s:CCsearches_colored}, but with 
considerably less complication.

\subsection{White signal+noise, stationary data}
\label{s:white}

To start, let us consider two coincident and coaligned 
detectors with uncorrelated noise.
We take both the detector noise and correlated stochastic 
signal to be Gaussian, stationary and white. 
We denote the variances by 
$\sigma_{n_1}^2$, $\sigma_{n_2}^2$, $\sigma_h^2$, respectively.
We will not assume that we know the noise variances a~priori,
so we will try to recover $\sigma^2_{n_1}$ and $\sigma^2_{n_2}$ in addition
to $\sigma^2_h$.

Then the likelihood function is given by
\be
p(d|\sigma^2_{n_1},\sigma^2_{n_2}, \sigma_h^2) 
=\frac{1}{\sqrt{\det(2\pi C)}}\, e^{-\frac{1}{2} d^T C^{-1} d}\,,
\label{e:likelihood_white}
\ee
where
\be
C 
\equiv \left[
\begin{array}{cc}
(\sigma^2_{n_1} +\sigma^2_h)\,\mathds{1}_{N\times N} & \sigma^2_h\,\mathds{1}_{N\times N} 
\\
\sigma^2_h\,\mathds{1}_{N\times N} & (\sigma^2_{n_2} +\sigma^2_h)\,\mathds{1}_{N\times N}
\\
\end{array}
\right]
\label{e:C_white}
\ee
is the covariance matrix and
\be
d^T C^{-1} d
\equiv \sum_{\alpha,\beta} \sum_{i,j}
d_{\alpha i} \left(C^{-1}\right)_{\alpha i,\beta j} d_{\beta j}\,.
\label{e:argexp_white}
\ee
The indices $i, j=1,2,\cdots, N$ label individual time samples, and 
$\alpha, \beta=1,2$ label the two detectors.
The joint posterior distribution for the signal and noise variances is
\begin{multline}
p(\sigma^2_{n_1}, \sigma^2_{n_2},\sigma^2_h|d) 
=\frac{p(d|\sigma^2_{n_1}, \sigma^2_{n_2}, \sigma_h^2)\> 
p(\sigma^2_{n_1}, \sigma^2_{n_2}, \sigma_h^2)}{p(d)}\,,
\label{e:posterior}
\end{multline}
where $p(\sigma^2_{n_1}, \sigma^2_{n_2}, \sigma_h^2)$ is the joint prior probability
distribution, and 
\begin{multline}
{p(d)}=
\int {\rm d} \sigma^2_{n_1}\int {\rm d} \sigma^2_{n_2} \int {\rm d} \sigma^2_h\>
\\
\times 
p(d|\sigma^2_{n_1}, \sigma^2_{n_2}, \sigma_h^2) 
p(\sigma^2_{n_1}, \sigma^2_{n_2}, \sigma_h^2)
\end{multline}
is the evidence for this signal+noise model.

It is easy to show that the maximum-likelihood estimators for the
parameters $\sigma^2_{n_1}$, $\sigma^2_{n_2}$, $\sigma_h^2$ are 
given by the following quadratic combinations of the data:
\be
\begin{aligned}
&\hat \sigma^2_{n_1} \equiv \frac{1}{N}\sum_i d_{1i}^2 
- \frac{1}{N}\sum_i d_{1i} d_{2i}\,,
\\
&\hat \sigma^2_{n_2} \equiv \frac{1}{N}\sum_i d_{2i}^2 
- \frac{1}{N}\sum_i d_{1i} d_{2i}\,,
\\
&\hat \sigma^2_h \equiv \frac{1}{N}\sum_i d_{1i} d_{2i}\,.
\end{aligned}
\label{e:MLestimators_white}
\ee
These expressions show that it is also convenient to define 
\be
\hat \sigma^2_{1} \equiv \frac{1}{N}\sum_i d_{1i}^2\,,
\qquad
\hat \sigma^2_{2} \equiv \frac{1}{N}\sum_i d_{2i}^2\,,
\label{e:S1S2_white}
\ee
which are estimators of the {\tt total} auto-correlated 
variances in the two detectors:
\be
\sigma^2_1\equiv \sigma^2_{n_1}+ \sigma^2_h\,,
\qquad 
\sigma^2_2\equiv \sigma^2_{n_2}+ \sigma^2_h\,.
\label{e:S1S2_def}
\ee
In the weak-signal limit $\sigma^2_1\approx \sigma^2_{n_1}$ and
$\sigma^2_2\approx \sigma^2_{n_2}$, but we will {\it not} make that 
approximation at this stage.
We will discuss this approximation in 
Sec.~\ref{s:white_reduced}.

To show that the above estimators are sufficient 
statistics for this problem, 
it suffices to show that the data enter the likelihood function 
\eqref{e:likelihood_white}  
{\it solely} in the form of these estimators, 
up to an overall normalization that does not depend on the 
signal and noise parameters.
Since the only part of the likelihood function that depends
on the data is the argument of the exponential,
we just need to show that $d^T C^{-1} d$ can be written in
terms of 
$\hat \sigma_{n_1}^2$, $\hat \sigma_{n_2}^2$, $\hat \sigma_h^2$, 
or, equivalently, in terms of 
$\hat \sigma_1^2$, $\hat \sigma_2^2$, $\hat \sigma_h^2$. 

Since the covariance matrix $C$ is a $2\times 2$-block matrix 
with each block proportional to $\mathds{1}_{N\times N}$, 
we can explicitly invert $C$, yielding
\be
C^{-1}
%&= \frac{1}{\beta}\left[
%\begin{array}{cc}
%(\sigma^2_{n_2} +\sigma^2_h)\,\mathds{1}_{N\times N} & -\sigma^2_h\,\mathds{1}_{N\times N} 
%\\
%-\sigma^2_h\,\mathds{1}_{N\times N} & (\sigma^2_{n_1} +\sigma^2_h)\,\mathds{1}_{N\times N}
%\\
%\end{array}
%\right]
%\\
= \frac{1}{\beta}\left[
\begin{array}{cc}
\sigma^2_2\,\mathds{1}_{N\times N} & -\sigma^2_h\,\mathds{1}_{N\times N} 
\\
-\sigma^2_h\,\mathds{1}_{N\times N} & \sigma^2_1\,\mathds{1}_{N\times N}
\\
\end{array}
\right]\,,
\label{e:Cinv_white}
\ee
where 
\be
\begin{aligned}
\beta
&\equiv \sigma^2_1 \sigma^2_2 - (\sigma^2_h)^2
\\
&= \sigma_{n_1}^2 \sigma_{n_2}^2 + \left(\sigma_{n_1}^2+\sigma_{n_2}^2 \right) \sigma_h^2  > 0\,.
\end{aligned}
\ee
Using this result, it is then straightforward to show that
\be
\begin{aligned}
d^T C^{-1} d 
&= \frac{1}{\beta}
\left[
\sigma^2_2\sum_i d_{1i}^2+\sigma^2_1\sum_i d_{2i}^2
-2\sigma^2_h\sum_i d_{1i} d_{2i}\right]
\\
%&= \frac{N}{\beta}
%\left[\sigma^2_2\hat \sigma^2_1 + \sigma^2_1\hat \sigma^2_2 -2\sigma^2_h\hat \sigma^2_h\right]
%\\
&= \frac{N}{(1-(\sigma^2_h)^2/\sigma^2_1 \sigma^2_2)}\left[
\frac{\hat \sigma^2_1}{\sigma^2_1} + \frac{\hat \sigma^2_2}{\sigma^2_2}
- 2 \frac{\sigma^2_h \hat \sigma^2_h}{\sigma^2_1 \sigma^2_2}\right]\,.
\label{e:dCinvd_white}
\end{aligned}
\ee
Thus,
\begin{multline}
p(d|\sigma^2_{n_1}, \sigma^2_{n_2}, \sigma_h^2)
=\frac{1}{(2\pi)^N (\sigma^2_1 \sigma^2_2 - (\sigma^2_h)^2)^{N/2}}
\\
\times\exp\left\{-\frac{1}{2}
\frac{N}{(1-(\sigma^2_h)^2/\sigma^2_1 \sigma^2_2)}\left[
\frac{\hat \sigma^2_1}{\sigma^2_1} + \frac{\hat \sigma^2_2}{\sigma^2_2}
- 2 \frac{\sigma^2_h \hat \sigma^2_h}{\sigma^2_1 \sigma^2_2}\right]\right\}\,,
\label{e:likefinal_white}
\end{multline}
which depends on the data only via
$\hat \sigma^2_1$, $\hat \sigma^2_2$, $\hat \sigma^2_h$, 
or, equivalently,
$\hat \sigma^2_{n_1}$, $\hat \sigma^2_{n_2}$, $\hat \sigma^2_h$. 

%%%%%%%%%%%%%%%%%%%%%%%%%%%%%%%%%%%%%%%%%%%%%%%%%%%%%%%%%
\subsection{Weak-signal approximation, estimating detector auto-correlations}
\label{s:white_reduced}

We now consider a {\it reduced} signal+noise model and its corresponding 
likelihood function, again for the case of white signal+noise, 
stationary data.
The reduction in the model has two components:
First, instead of treating the detector noise variances 
$\sigma^2_{n_1}$, $\sigma^2_{n_2}$ or the total auto-correlated variances
$\sigma^2_1$, $\sigma^2_2$ as free parameters for our analysis, we will 
use {\it measured estimates} $\bar \sigma^2_1$, $\bar \sigma^2_2$ in 
place of $\sigma^2_1$, $\sigma^2_2$.
We use overbars instead of hats to denote these 
quantities to indicate that $\bar \sigma^2_1$, $\bar \sigma^2_2$ are 
not the same as the maximum-likelihood data combinations 
$\hat \sigma^2_1$, $\hat \sigma^2_2$ for the segment that we are analyzing.
For example, for the actual LIGO-Virgo stochastic searches, 
the estimates of the auto-correlated power are 
constructed from two segments of data (each approximately one minute in duration) 
immediately preceding and following the analysis segment.
In fact, we will show in Fig.~\ref{fig:pp_plot_white} that if the
measured estimates of the autocorrelated power equal the 
maximum-likelihood data combinations for the analysis segment, then we 
obtain biased recoveries of the GWB spectrum, due to covariances between 
the autocorrelation and cross-correlation estimators~\cite{Lazzarini:bias}.
(But see also the discussion of bias 
at the end of Appendix~\ref{s:alternative_derivation}.)

Second, we will assume that the stochastic signal is {\it weak}
compared to the detector noise and thus keep only the 
leading-order terms in expressions involving 
$(\sigma^2_h)^2/\bar \sigma^2_1\bar \sigma^2_2\ll 1$.
To zeroth order, the detector noise and total auto-correlated
variances are equal to one another, 
$\sigma^2_{n_1}\approx \sigma^2_1$, 
$\sigma^2_{n_2}\approx \sigma^2_2$.
This weak-signal approximation is valid for searches for GWBs 
using ground-based detectors like LIGO and Virgo.
It is not a good approximation for GWB searches using pulsar
timing arrays, where the auto-correlated power in the GWB may 
exceed that due
to pulsar noise and timing measurement noise at very low 
frequencies~\cite{Siemens-et-al:2013, arzoumanian2020nanograv}.

The likelihood function for the reduced signal+noise model 
can be obtained from \eqref{e:likefinal_white} by making the 
simplifications described above.
We first approximate the terms $\hat\sigma_1^2/\sigma_1^2$ and 
$\hat\sigma_2^2/\sigma_2^2$ by 1, given that we are 
replacing the parameters $\sigma_1^2$ and $\sigma_2^2$ by 
measured estimates of these quantities.
We then replace the remaining factors of 
$\sigma_1^2$, $\sigma_2^2$, which appear in 
lihood 
function in the combination $1/\sigma_1^2\sigma_2^2$ by 
\be
\frac{1}{\bar\Sigma_{12}^4}\equiv
\frac{1}{\bar\sigma_1^2\bar\sigma_2^2(1+2/N_{\rm avg})}\,,
\label{e:unbiased_1/P2}
\ee
where $N_{\rm avg}$ (which we assume to be equal for both
detectors) is the number of averages used in the construction 
of $\bar\sigma_1^2$, $\bar\sigma_2^2$, e.g., Welch power
spectrum estimates~\cite{Welch:1967}.
The justification for including the factor of $(1+2/N_{\rm avg})$
is given in Appendix~\ref{s:psd_uncertainties}; the factor 
removes a bias that would otherwise exist in the estimation of 
$1/\sigma_1^2\sigma_2^2$, due to the use of 
a finite amount of data to construct $\bar\sigma_1^2$, $\bar\sigma_2^2$.
Making all these replacments in \eqref{e:likefinal_white}, 
we obtain
\begin{widetext}
\be
p(d|\bar \sigma^2_1, \bar \sigma^2_2, \sigma_h^2)
=\frac{1}{(2\pi)^N (\bar \Sigma_{12}^4)^{N/2} 
(1- (\sigma^2_h)^2/\bar \Sigma_{12}^4)^{N/2}}
\exp\left\{-\frac{N}{(1-(\sigma^2_h)^2/\bar \Sigma_{12}^4)}
\left[1 - \frac{\sigma^2_h \hat \sigma^2_h}{\bar \Sigma_{12}^4}\right]\right\}\,.
\label{e:temp1}
\ee

We then Taylor expand the RHS keeping only the leading-order 
terms in 
$(\sigma^2_h)^2/\sigma_1^2\sigma_2^2 = (\sigma^2_h)^2/\bar \Sigma_{12}^4$.
The $\sigma^2_h$ factor in front of the exponential can be written as
\be
\frac{1}{(1-(\sigma^2_h)^2/\bar \Sigma^4_{12})^{N/2}}
\approx 1+\frac{N}{2}\frac{(\sigma^2_h)^2}{\bar \Sigma^4_{12}}
\approx\exp\left[\frac{N}{2}\frac{(\sigma^2_h)^2}{\bar \Sigma^4_{12}}\right]\,,
\ee
while the exponential factor itself becomes
\be
\exp\left\{-\frac{N}{(1-(\sigma^2_h)^2/\bar \Sigma^4_{12})}
\left[1 - \frac{\sigma^2_h \hat \sigma^2_h}{\bar \Sigma^4_{12}}\right]\right\}
\approx
\exp\left\{-{N}\left(1+\frac{(\sigma^2_h)^2}{\bar \Sigma^4_{12}}
- \frac{\sigma^2_h \hat \sigma^2_h}{\bar \Sigma^4_{12}}\right)\right\}\,.
\ee
Combining these last two expressions gives
\be
\exp\left[-\frac{N}{2}\left(2 + 
\frac{(\sigma^2_h)^2-2 \sigma^2_h\hat \sigma^2_h}{\bar \Sigma^4_{12}}\right)\right]
=
e^{-N}
\exp\left[\frac{(\hat \sigma^2_h)^2}{2\,{\rm var}(\hat\sigma^2_h)}\right]
\exp\left[-\frac{(\hat \sigma^2_h-\sigma^2_h)^2}{2\,{\rm var}(\hat\sigma^2_h)}\right]\,,
\ee
where we completed the square in $\sigma^2_h$ and $\hat \sigma^2_h$ and made the substitution
\be
{\rm var}({\hat \sigma^2_h}) 
\equiv {\bar \Sigma^4_{12}}/{N}=
\left(1 + {2}/{N_{\rm avg}}\right){{\bar \sigma^2_1\bar \sigma^2_2}}/{N}\,,
\label{e:reduced_var_white}
\ee
which is the leading-order expression for the variance of the maximum-likelihood
estimator $\hat \sigma^2_h$.
Thus, 
\be
p(d|\bar \sigma_1^2, \bar \sigma_2^2, \sigma_h^2)
=\frac{e^{-N}}{(2\pi)^N (N{\rm var}(\hat\sigma_h^2))^{N/2}} 
\exp\left[\frac{(\hat \sigma^2_h)^2}{2\,{\rm var}(\hat\sigma^2_h)}\right]
\exp\left[-\frac{(\hat \sigma^2_h-\sigma^2_h)^2}{2\,{\rm var}(\hat\sigma^2_h)}\right]\,,
\label{e:likefinal_white_reduced}
\ee
which shows that $\hat \sigma^2_h$ and ${\rm var}(\hat\sigma^2_h)$
are sufficient statistics for this reduced signal+noise model.
Note that the $\sigma^2_h$-dependent term in the likelihood function has the same 
general form as that for the simple example described in Sec.~\ref{s:example};
see \eqref{e:p(d|a)_suffstat}.
(In Appendix~\ref{s:alternative_derivation}, we present an alternative 
derivation of the likelihood function for a reduced signal+noise model, 
but under different assumptions than used here.)
\end{widetext}

\begin{figure}[h!tbp]
\centering
\includegraphics[width=0.45\textwidth]{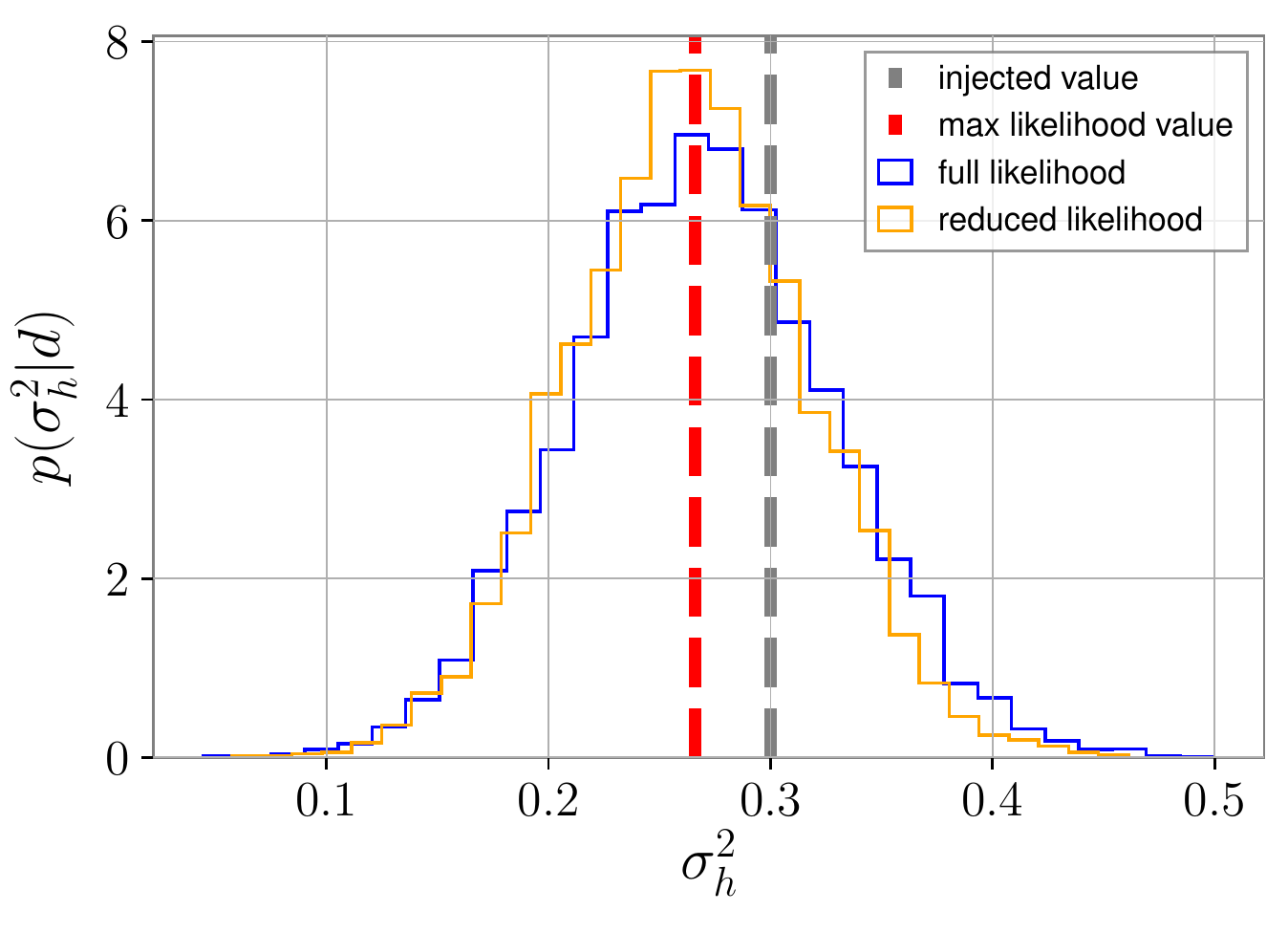}
\caption{Comparison of the marginalized posteriors for the full
likelihood function \eqref{e:likefinal_white} (blue histogram) 
and its reduced version \eqref{e:likefinal_white_reduced} (orange histogram), 
which substitutes
measured estimates $\bar \sigma^2_1$, $\bar \sigma^2_2$ for the auto-correlated
power $\sigma^2_1$, $\sigma^2_2$ and keeps only leading-order terms in 
$(\sigma^2_h)^2/\bar \sigma^2_1\bar \sigma^2_2$.}
\label{f:comparison_suffstat_reduced}
\end{figure}

Figure~\ref{f:comparison_suffstat_reduced} compares the marginalized
posterior distributions for $\sigma^2_h$ calculated from the two likelihood
functions, \eqref{e:likefinal_white} and \eqref{e:likefinal_white_reduced},
for a short stretch of simulated time-series data consisting of a white signal 
injected into white noise in two coincident and coaligned detectors.
The simulated noise had variances $\sigma^2_{n_1}=\sigma^2_{n_2}=1$, while
the injected signal had $\sigma^2_h=0.3$.
For 512~samples, these values correspond to an expected 
signal-to-noise ratio 
$\rho = \sqrt{N}\,{\sigma^2_h}/\sqrt{\sigma^2_1\sigma^2_2}=5.22$
for the full set of data.
Note, however, that $(\sigma^2_h)^2/\sigma^2_1\sigma^2_2 =0.05\ll 1$, 
consistent with the weak-signal approximation.
The blue histogram is the marginalized posterior for $\sigma^2_h$ calculated
from the full likelihood function \eqref{e:likefinal_white}; the orange histogram 
is the marginalized posterior for $\sigma^2_h$ calculated from the reduced 
likelihood function \eqref{e:likefinal_white_reduced}.
For the reduced likelihood analysis, the detector auto-correlated power were
estimated from an additional simulated data segment.  
The dashed vertical grey line and the dashed vertical red line show
the injected value of $\sigma^2_h$ and the maximum-likelihood value $\hat \sigma^2_h$.

%% file: crosscorr_colored.tex
\section{Sufficient statistics for cross-correlation searches
-- colored signal+noise models}
\label{s:CCsearches_colored}

Here we extend the analysis of the previous section to 
a colored signal+noise model, starting with stationary 
data in Sec.~\ref{s:colored_stationary} and then 
discussing the complications introduced by non-stationary 
noise in Sec.~\ref{s:colored_nonstationary}.
In Sec.~\ref{s:colored_reduced}, we simplify the 
signal+noise model by again considering weak signals 
and measured estimates of the auto-correlated power,
for which the frequency integrands of the
cross-correlation statistic and its variance
are sufficient statistics.
In Sec.~\ref{s:hybrid} we show that the 
full Bayesian analysis is approximately equivalent to
LIGO-Virgo's hybrid frequentist-Bayesian analysis, and
in Sec.~\ref{s:comparison} we construct percentile-percentile
(pp) plots~\cite{pp_plot} to show that the reduced analyses
have proper statistical coverage.
(We do not assume that the detectors are coincident and 
co-aligned in this section.)

%%%%%%%%%%%%%%%%%%%%%%%%%%%%%%%%%%%%%%%%%%%%%%%%%%%%%%%%%
\subsection{Colored signal+noise, stationary data}
\label{s:colored_stationary}

For the case where the detector noise and GWB signal are 
colored, it simplest to work in the frequency domain,
since the Fourier components are {\it independent} of one
another (assuming the data are stationary over the 
duration of the analysis segment).
Assuming multivariate Gaussian distributions as before,
the variances 
$\sigma^2_{n_1}$, $\sigma^2_{n_2}$, $\sigma^2_h$, 
for the white signal+noise
case are replaced by
{\it power spectral densitites} 
$P_{n_1}(f)$, $P_{n_2}(f)$, $P_h(f)$, 
defined by
\be
\langle \tilde h(f)\tilde h^*(f')\rangle
= \frac{1}{2}\delta(f-f') P_h(f)\,,
\quad{\rm etc.},
\ee
where $\tilde h(f)$ is the Fourier transform of the 
signal component $h(t)$ of the time-series data.
The factor of $1/2$ is included to make these {\it one-sided}
power spectral densities, so that the total variance is 
given by an integral of the power spectral density over
{\it positive frequencies}, 
$\sigma^2 = \int_0^\infty {\rm d}f\> P(f)$.
Although the above expressions are written in terms of
continuous functions of frequency, in practice we work
with frequency series, e.g., $P_h(f_k)$, where the 
discrete frequencies take values $f_k \equiv k\Delta f$, 
where $\Delta f \equiv 1/T$ and $k=0, 1, \cdots, N/2 -1$, for 
a data segment of duration $T\equiv N\Delta t$.
For white data, $P(f) = 2\sigma^2\Delta t$ is constant between
$f=0$ and the Nyquist frequency $f_{\rm Nyq} \equiv 1/2\Delta t$.

Even though we are assuming here that the data are stationary, 
it is convenient to divide the data from a large observation 
period into segments, which we will label by $I=1,2,\cdots, N_{\rm seg}$.
The full likelihood function is then a product of the 
likelihood functions for the individual segments
\be
p(d|P_{n_1}, P_{n_2}, P_h)
= \prod_I p_I(d_I|P_{n_1}, P_{n_2}, P_h)\,,
\label{e:product-likelihood-colored-segment}
\ee
where
\begin{multline}
p_I(d_I|P_{n_1}, P_{n_2}, P_h)
\\
= \prod_{k} \frac{1}{{\rm det}(2\pi \tilde C(f_k))}
e^{-\frac{1}{2}\sum_{\alpha,\beta} 
\tilde d_{\alpha I}^*(f_k) 
\left(\tilde C(f_k)^{-1}\right)_{\alpha\beta}
\tilde d_{\beta I}(f_k)}\,.
\label{e:likelihood-colored-segment}
\end{multline}
In the above expression
\be
\tilde C(f)
= \frac{T}{4}\left[
\begin{array}{cc}
P_1(f) & \gamma(f)P_h(f)
\\
\gamma(f)P_h(f) & P_2(f)
\end{array}
\right]
\label{e:Ctilde}
\ee
is the covariance matrix of the data, with
\be
\begin{aligned}
&P_1(f)\equiv P_{n_1}(f) + P_h(f)\,,
\\
&P_2(f)\equiv P_{n_2}(f) + P_h(f)\,,
\end{aligned}
\ee
denoting the total auto-correlated detector
power spectral densities.
The inverse covariance matrix is simply
\begin{multline}
\tilde C(f)^{-1}
=\frac{4}{T}\frac{1}{(P_{1}(f)P_{2}(f) - \gamma^2(f)P_h^2(f))}
\\
\times\left[
\begin{array}{cc}
P_{2}(f) & -\gamma(f)P_h(f)
\\
-\gamma(f)P_h(f) & P_{1}(f)
\end{array}
\right]\,.
\end{multline}
Note that although the data depend on segment $I$, 
the parameters $P_{n_1}$, $P_{n_2}$, $P_h$ 
do not, since we are assuming that the noise and 
signal power spectra are the same in each segment.

The dimensionless function $\gamma(f)$, which appears 
in the off-diagonal elements of the covariance matrix, 
is the {\it overlap reduction function}, which
accounts for the relative position and orientation of the detectors
\cite{christensen92,Flanagan:1993}. The functional form of $\gamma(f)$ is not
relevant for the discussion that follows, other than the fact that $\gamma(f)$
equals unity for coincident and coaligned detectors in the long-wavelength limit.

If we want to estimate the values of the power 
spectral densities 
$P_{n_1}(f)$, $P_{n_2}(f)$, $P_h(f)$, at 
{\it each} discrete (positive) frequency $f_k\equiv k\Delta f$,
then there are no simplying 
sufficient statistics for this case as the data enter the 
likelihood function only through the combinations
\be
|\tilde d_{1I}(f_k)|^2\,,
\quad
|\tilde d_{2I}(f_k)|^2\,,
\quad
{\rm Re} \left(\tilde d_{1I}^*(f_k)\tilde d_{2I}(f_k) \right)\,,
\label{e:single-freq-estimators}
\ee
which does not correspond to a reduction in the number of 
data samples used in writing the likelihood function.
However, if the power spectra are expected to be smooth
over a {\it coarser} frequency resolution
$\delta f \equiv 1/\tau > \Delta f$, 
where $\tau\equiv T/M$ is some 
fractional part of the segment duration $T$, then there
is a reduction in the data combinations.
This is because the 
relevant power spectra need only be estimated at fewer
discrete frequencies $f_\ell \equiv\ell\,\delta f$, 
where $\ell = 0, 1,\cdots, (N/M)/2-1$.
(For typical LIGO-Virgo searches, $M$ is approximately 20.)
Hence the data combinations
\eqref{e:single-freq-estimators} can be averaged
over a subset of $M$ fine-grained frequencies $f_k$ centered at
$f_\ell$:
\be
\begin{aligned}
\tilde c_{11,I}(f_\ell)
&\equiv \frac{1}{M}\sum_{k=\ell-M/2}^{\ell+M/2-1}
|\tilde d_{1I}(f_k)|^2\,,
\\
\tilde c_{22,I}(f_\ell)
&\equiv \frac{1}{M}\sum_{k=\ell-M/2}^{\ell+M/2-1}
|\tilde d_{2I}(f_k)|^2\,,
\\
\tilde c_{12,I}(f_\ell)
&\equiv \frac{1}{M}\sum_{k=\ell-M/2}^{\ell+M/2-1}
{\rm Re} \left(\tilde d_{1I}^*(f_k) \tilde d_{2I}(f_k)\right)\,.
\end{aligned}
\label{e:coarse-grained-data}
\ee
This leads to averaged (or {\it coarse-grained}) 
power spectral density estimators 
\be
\begin{aligned}
&\hat P_{1I}(f_\ell)
\equiv \frac{2}{T}\,\tilde c_{11,I}(f_\ell)\,,
\\
&\hat P_{2I}(f_\ell)
\equiv \frac{2}{T}\,\tilde c_{22,I}(f_\ell)\,,
\\
&\hat P_{hI}(f_\ell)
\equiv \frac{2}{T}\,\frac{\tilde c_{12,I}(f_\ell)}{\gamma(f_\ell)}\,,
\label{e:coarse-grained-estimators}
\end{aligned}
\ee
in terms of which the likelihood function \eqref{e:likelihood-colored-segment}
can be written:
\begin{widetext}
\begin{multline}
p_I(d_I|P_{n_1}, P_{n_2}, P_h)
=\prod_\ell
\frac{1}{(\pi T/2)^{2M}
(P_1(f_\ell)P_2(f_\ell) - \gamma^2(f_\ell)P_h^2(f_\ell))^{M}}
\\
\times
\exp\left\{-\frac{M}
{(1-\gamma^2(f_\ell)P_h^2(f_\ell)/P_1(f_\ell)P_2(f_\ell))}\left[
\frac{\hat P_{1I}(f_\ell)}{P_1(f_\ell)} + \frac{\hat P_{2I}(f_\ell)}{P_2(f_\ell)}
- 2 \gamma^2(f_\ell) \frac{P_h(f_\ell) \hat P_{hI}(f_\ell)}{P_1(f_\ell)P_2(f_\ell)}
\right]\right\}\,.
\label{e:likefinal_colored_coarsegrained}
\end{multline}
We note that estimating a power spectrum by subdividing a 
segment of data into shorter duration subsegments is a 
standard practice in signal processing~\cite{Welch:1967}.
Its a way to reduce the variance in the power spectrum
estimate at the expense of a coarser frequency resolution.
\end{widetext}

For this stationary case, we get a further level of data reduction
in the sufficient statistics, 
as we can average the coarsed-grained power spectrum 
estimators \eqref{e:coarse-grained-estimators} over the
number of segments.
%
%\be
%\begin{aligned}
%&\hat P_{h}(f_\ell)
%\equiv \frac{1}{N_{\rm seg}}\sum_I
%\frac{2}{T}\,\frac{\tilde c_{12,I}(f_\ell)}{\gamma(f_\ell)}\,,
%\\
%&\hat P_{1}(f_\ell)
%\equiv \frac{1}{N_{\rm seg}}\sum_I
%\frac{2}{T}\,\tilde c_{11,I}(f_\ell)\,,
%\\
%&\hat P_{2}(f_\ell)
%\equiv \frac{1}{N_{\rm seg}}\sum_I
%\frac{2}{T}\,\tilde c_{22,I}(f_\ell)\,.
%\label{e:coarse-grained-estimators-stationary}
%\end{aligned}
%\ee
%
By construction, these segment-averaged estimators will 
give expected values of 
$P_h(f_\ell)$, $P_1(f_\ell)$, $P_2(f_\ell)$ 
over the {\it whole} observation.
This is fine for stationary data. 
But if the detector 
noise levels can change from segment to segment, then 
this simple averaging will fail to capture the 
non-stationarity of the noise.
We have to do something different for this more complicated,
but realistic, scenario.

%%%%%%%%%%%%%%%%%%%%%%%%%%%%%%%%%%%%%%%%%%%%%%%%%%%%%%%%%
\subsection{Colored signal+noise, non-stationary data}
\label{s:colored_nonstationary}

For non-stationary detector noise, we have to increase the number 
of model parameters from 
$P_{n_1}(f_\ell)$, $P_{n_2}(f_\ell)$, $P_h(f_\ell)$ to
$P_{n_1I}(f_\ell)$, $P_{n_2I}(f_\ell)$, $P_h(f_\ell)$, 
where $I=1,2,\cdots, N_{\rm seg}$, since the noise levels 
can differ from segment to segment.
We are assuming here that the power spectrum of the 
stochastic signal is stationary, which is not necessarily 
the case for a ``popcorn-like" background, such as that
produced by stellar-mass binary black hole mergers~\cite{tbs_methods}. 
The covariance matrix for this case is then
\be
\tilde C_I(f_\ell)
=\frac{T}{4}
\left[
\begin{array}{cc}
P_{1I}(f_\ell) & \gamma(f_\ell) P_h(f_\ell)
\\
\gamma(f_\ell) P_h(f_\ell) & P_{2I}(f_\ell)
\end{array}
\right]\,,
\ee
where
\be
\begin{aligned} 
&P_{1I}(f)\equiv P_{n_1 I}(f) + P_h(f)\,,
\\
&P_{2I}(f)\equiv P_{n_2 I}(f) + P_h(f)\,,
\end{aligned}
\ee
are the auto-correlated detector power spectra 
for segment $I$.
Similar to what we found in
\eqref{e:likefinal_colored_coarsegrained}, the 
corresponding likelihood function 
for a single segment of data can be written as
\begin{widetext}
\begin{multline}
p_I(d_I|P_{n_1 I}, P_{n_2 I}, P_h)
=\prod_\ell
\frac{1}{(\pi T/2)^{2M}
(P_{1I}(f_\ell)P_{2I}(f_\ell) - \gamma^2(f_\ell)P_h^2(f_\ell))^{M}}
\\
\times
\exp\left\{-\frac{M}
{(1-\gamma^2(f_\ell)P_h^2(f_\ell)/P_{1I}(f_\ell)P_{2I}(f_\ell))}\left[
\frac{\hat P_{1I}(f_\ell)}{P_{1I}(f_\ell)} + \frac{\hat P_{2I}(f_\ell)}{P_{2I}(f_\ell)}
- 2 \gamma^2(f_\ell) \frac{P_h(f_\ell) \hat P_{hI}(f_\ell)}{P_{1I}(f_\ell)P_{2I}(f_\ell)}
\right]\right\}\,,
\label{e:likefinal_colored_nonstationary}
\end{multline}
where the estimators
$\hat P_{hI}(f_\ell)$, $\hat P_{1I}(f_\ell)$, $\hat P_{2I}(f_\ell)$ are
the same as in \eqref{e:coarse-grained-estimators}.
We emphasize that this likelihood differs from that in 
\eqref{e:likefinal_colored_coarsegrained} only by our 
assumption that the noise is not stationary, which is 
reflected in the fact that the parameters $P_{1I}(f_\ell)$, $P_{2I}(f_\ell)$ 
carry $I$ indices.
Note that the above likelihood has the same form as the white noise case
\eqref{e:likefinal_white}, but with both frequency and segment dependence 
in the estimators and parameters.
The full expression for the likelihood function involves 
a further product over $I$, as specified in 
\eqref{e:product-likelihood-colored-segment}.

%%%%%%%%%%%%%%%%%%%%%%%%%%%%%%%%%%%%%%%%%%%%

\subsection{Reduced version of the colored, non-stationary 
likelihood function}
\label{s:colored_reduced}

To simplify the above analysis, we will proceed as we did
in Sec.~\ref{s:white_reduced}, where we considered a reduced 
signal+noise model and its corresponding likelihood by 
replacing the auto-correlated power spectral densities
with measured estimates $\bar P_{1I}(f_\ell)$, $\bar P_{2I}(f_\ell)$,
and working in the weak-signal approximation where
$P^2_h(f_\ell)/\bar P_{1I}(f_\ell) \bar P_{2I}(f_\ell)\ll 1$.
For  a given discrete frequency $f_\ell$ and data segment $I$,
the reduction in the likelihood function has exactly the same 
form as the white noise case, 
which allows us to immediately write down:
\be
p_I(d_I|\bar P_{1I}, \bar P_{2I}, P_h)
=\prod_\ell
\frac{e^{-2M}}
{(\pi T/2)^{2M}(\bar P_{1I}(f_\ell) \bar P_{2I}(f_\ell)(1+2/N_{\rm avg}))^M}
\exp\left[\frac{\hat P_{hI}^2(f_\ell)}{2\bar\sigma^2_{hI}(f_\ell)}\right]
\exp\left[-\frac{(\hat P_{hI}(f_\ell)-P_h(f_\ell))^2}{2\bar\sigma^2_{hI}(f_\ell)}\right]\,,
\ee
where
\be
\bar \sigma^2_{hI}(f_\ell) \equiv \frac{1}{2T\delta f}
\frac{\bar P_{1I}(f_\ell)\bar P_{2I}(f_\ell)}{\gamma^2(f_\ell)}
(1+2/N_{\rm avg})\,,
\label{e:reduced_var_colored}
\ee
which is the leading-order expression for the variance of $\hat P_{hI}(f_\ell)$.
(Here we used the relation $M=T\delta f$ for the coarse-grained frequencies,
and we included the factor of $(1+2/N_{\rm avg})$ to account for imperfect
estimation of the detector auto-correlated power spectra $\bar  P_{1I}(f_{\ell})$,
$\bar P_{2I}(f_\ell)$.)
Additionally,
since the parameter $P_h(f_\ell)$ shows up only in the last exponential,
we can use identity \eqref{e:xopt_identity} from Appendix~\ref{s:useful_identity}
to perform the product over the number of data segments, which translates 
into a sum over $I$ inside the exponential:
\be
\sum_I \frac{(\hat P_{hI}(f_\ell)-P_h(f_\ell))^2}{\bar\sigma^2_{hI}(f_\ell)}
= \sum_I \frac{\hat P_{hI}^2(f_\ell)}{\bar\sigma^2_{hI}(f_\ell)}
-\frac{\hat P_{h,{\rm opt}}^2(f_\ell)}{\bar\sigma_{h,{\rm opt}}^2(f_\ell)}
+\frac{(\hat P_{h,{\rm opt}}(f_\ell) -P_h(f_\ell))^2}{\bar\sigma^2_{h,{\rm opt}}(f_\ell)}\,,
\ee
where 
\begin{align}
\frac{\hat P_{h,{\rm opt}}(f_\ell)}{\bar\sigma^2_{h,{\rm opt}}(f_\ell)}
&\equiv\sum_I
\frac{2T\delta f\gamma(f_\ell)}{\bar P_{1I}(f_\ell)\bar P_{2I}(f_\ell)}
\frac{1}{(1+2/N_{\rm avg})}
%\nonumber\\
%&\hspace{1.25in}\times 
\frac{2}{T}\frac{1}{M}\sum_{k=\ell-M/2}^{\ell+M/2-1}
{\rm Re}\left(\tilde d_{1I}^*(f_k) \tilde d_{2I}(f_k)\right)\,,
\label{e:Sh_opt_freq}
\\
\frac{1}{\bar\sigma^2_{h,{\rm opt}}(f_\ell)}
&\equiv\sum_I
\frac{2T\delta f\gamma^2(f_\ell)}{\bar P_{1I}(f_\ell)\bar P_{2I}(f_\ell)}
\frac{1}{(1+2/N_{\rm avg})}\,.
\label{e:varh_opt_freq}
\end{align}
Thus, summing over both $I$ and $\ell$:
\begin{multline}
p(d|\{\bar P_{1I}\}, \{\bar P_{2I}\}, P_h)
\\
=\prod_\ell
\frac{e^{-2MN_{\rm seg}}}
{(\pi T/2)^{2MN_{\rm seg}}
\prod_I(\bar P_{1I}(f_\ell) \bar P_{2I}(f_\ell))^M}
\exp\left[\frac{\hat P_{h,{\rm opt}}^2(f_\ell)}{2\bar\sigma_{h,{\rm opt}}^2(f_\ell)}\right]
\exp\left[-\frac{(\hat P_{h,{\rm opt}}(f_\ell) -P_h(f_\ell))^2}{2\bar\sigma^2_{h,{\rm opt}}(f_\ell)}\right]\,.
\label{e:likefinal_colored_reduced}
\end{multline}
This is our main result.
Note that it has the same basic form as \eqref{e:likefinal_white_reduced},
which we derived for the white signal+noise case.
Thus, given some choice for the prior probability $p(P_h(f_\ell))$, 
the posterior distribution for $P_h(f_\ell)$ is given by
\be
p(P_h(f_\ell)|d, \{\bar P_{1I}(f_\ell)\}, \{\bar P_{2I}(f_\ell)\})
\propto 
\exp\left[-\frac{(\hat P_{h,{\rm opt}}(f_\ell) -P_h(f_\ell))^2}{2\bar\sigma^2_{h,{\rm opt}}(f_\ell)}\right]
\,p(P_h(f_\ell))\,.
\ee
This expression for the posterior shows that 
$\hat P_{h,{\rm opt}}(f_\ell)$ and $\bar\sigma_{h,{\rm opt}}^2(f_\ell)$ given by
\eqref{e:Sh_opt_freq} and \eqref{e:varh_opt_freq} are sufficient 
statistics for this 
colored signal+noise, non-stationary analysis, assuming weak signals
and measured estimates of the auto-correlated power in two detectors
for each data segment.
\end{widetext}

\begin{figure}%[htbp!]
\centering
\includegraphics[width=0.45\textwidth]{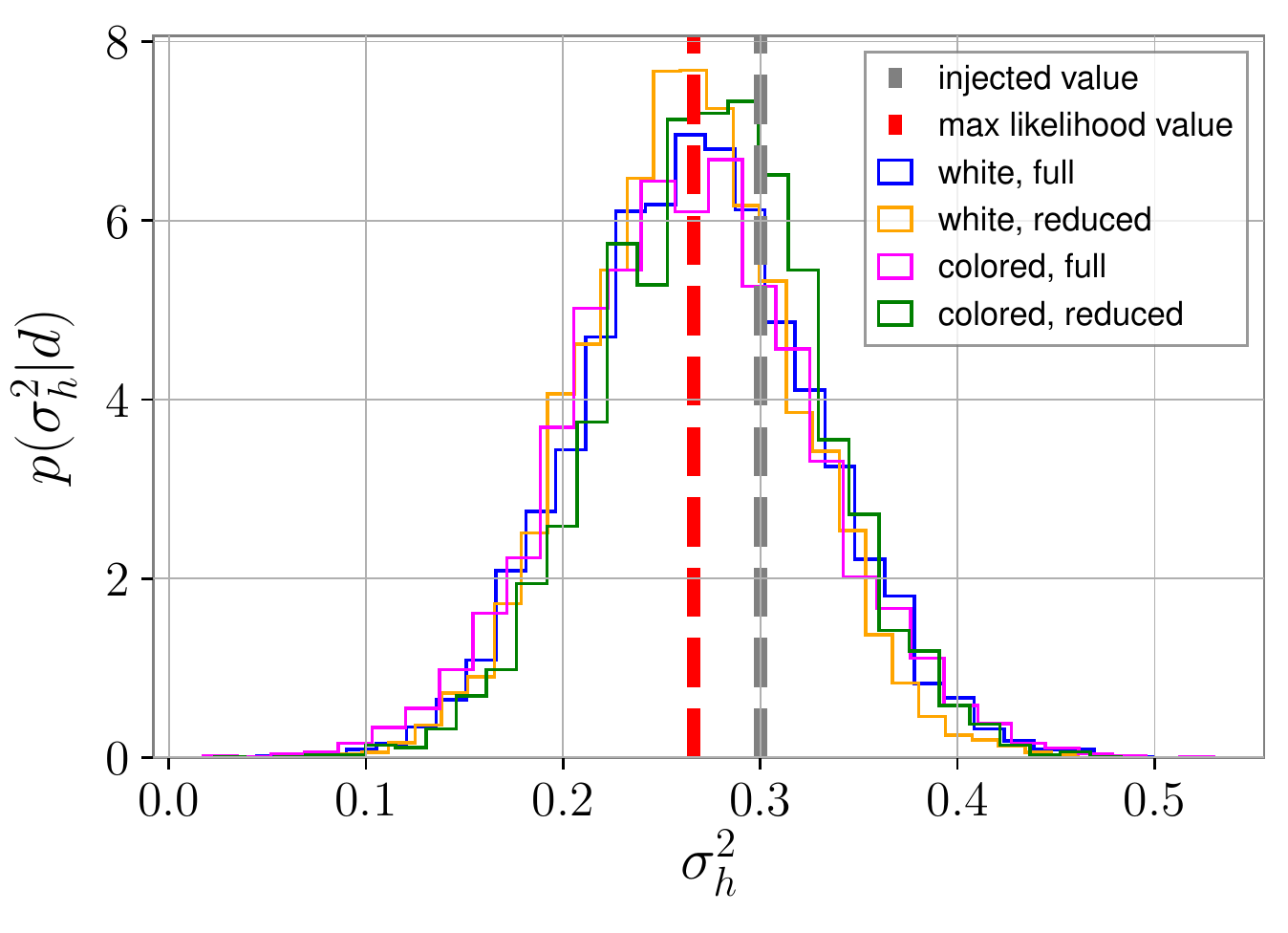}
\caption{Recovered posterior distributions for the amplitude of the 
GWB as obtained from the full and reduced versions of the Bayesian
likelihood functions appropriate for both white and colored data.
The simulated data consisted of a white signal and white detector noise.}
\label{fig:compareWhiteColoredReduced}
\end{figure}

In Figure~\ref{fig:compareWhiteColoredReduced}, we compare recoveries of 
the amplitude of the GWB using the full and reduced versions of the 
Bayesian likelihood functions appropriate for both white and colored data.
The simulated data are the same as that from the previous section, consisting
of a white GWB signal injected into white detector noise, for two coincident
and coaligned detectors.
For the reduced likelihood functions, the detector autocorrelated power were
estimated from an additional simulated data segment.
We see that all analyses agree very well, demonstrating that the 
mathematical derivations capture the behavior of the fully-Bayesian 
analysis in practice.

\begin{figure*}%[htbp!]
\centering
\includegraphics[width=\textwidth]{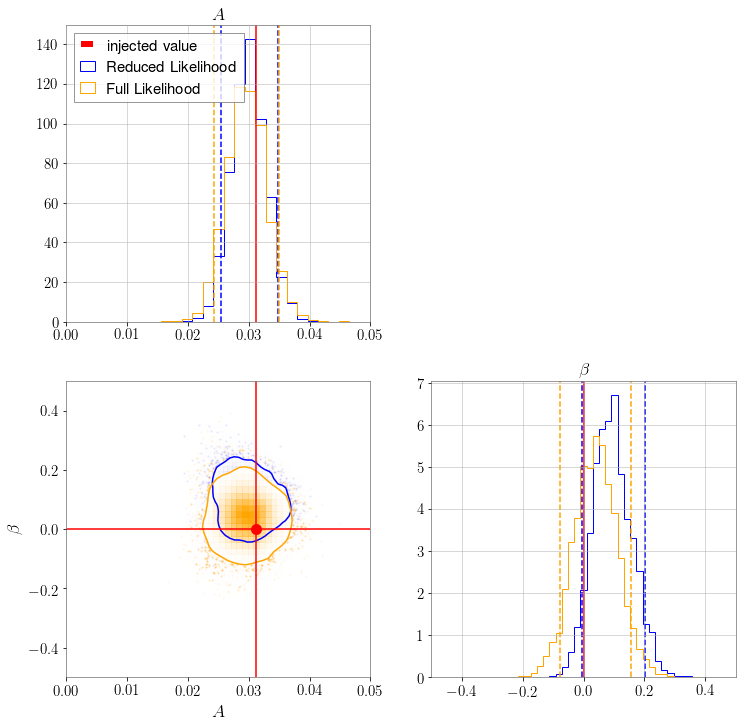}
%{comparePowerLawAmplitude}
%\includegraphics[width=0.45\textwidth]{comparePowerLawIndex}
\caption{Recovered posterior distributions obtained from 
the full and reduced likelihood functions for a colored signal+noise model.
The simulated data consisted of a white GWB signal injected into 
power-law detector noise.
The solid  red lines show the injected values of the GWB
amplitude and spectral index.}
\label{fig:ColoredInjectionComparison}
\end{figure*}

In Figure~\ref{fig:ColoredInjectionComparison}, we compare recoveries 
of the GWB amplitude and spectral index using the full and reduced
versions of the Bayesian likelihood function appropriate 
for a colored signal+noise model. 
For the full likelihood, we parameterize both the GWB signal and
detector noise as power laws of the form
\begin{equation}
P_h(f)= A \left(\frac{f}{f_{\rm ref}}\right)^\beta\,,
\qquad{\rm etc.}\,,
\end{equation}
where the $A$ is the amplitude, $\beta$ is the spectral index, 
and $f_{\rm ref}$ is a reference frequency.
(The amplitude and spectral indices for the detector noise
will differ, in general, from those for the GWB.)
For this simulation the injected noise power spectra have
$A_{n_1}=A_{n_2}=0.125$, and $\beta_1=\beta_2=0.5$,
while the injected signal has $A=A_{n_1}/2$, $\beta=0$ (i.e., it
is white).
For the reduced likelihood function, the detector 
auto-correlated 
power were estimated from coarse-grained power spectral density 
estimators applied to an additional simulated data segment.

Finally, we note that by using sufficient statistics, one
improves the computational efficiency of a search relative
to a fully-Bayesian analysis that works with the raw 
{\it uncombined} time-domain (or frequency-domain) data samples.
For example, 
for a stationary GWB, one can reduce the number of likelihood
evaluations by a factor of $MN_{\rm seg}$, where 
$M\sim 10$-20 corresponds to coarse-graining, and 
$N_{\rm seg}\sim 10^5$ comes from averaging six months of data 
split into approximately one-minute data segments.
One simply precomputes the sufficient statistic data combinations
\eqref{e:Sh_opt_freq} and \eqref{e:varh_opt_freq},
and performs Bayesian inference (e.g., MCMC sampling) for the 
corresponding likelihood function \eqref{e:likefinal_colored_reduced}.

\begin{widetext}
%

%%%%%%%%%%%%%%%%%%%%%%%%%%%%%%%%%%%%%%%%%%%
\subsection{Connection to LIGO's hybrid frequentist-Bayesian analysis}
\label{s:hybrid}

We can use the reduced-form of the likelihood 
\eqref{e:likefinal_colored_reduced}
to search for GWB signals described by the power spectral density $P_h(f)$.
But to make connection with the literature on GWBs, we should parametrize 
the background in terms of the (dimensionless) energy density spectrum~\cite{Allen_Romano_1999}:
\be
\Omega_{\rm gw}(f) \equiv
\frac{1}{\rho_c}\frac{{\rm d}\rho_{\rm gw}}{{\rm d}\ln f}\,,
\ee
where $\rho_c$ is the energy density needed to close the universe.
Then
\be
P_h(f) = \frac{3H_0^2}{10\pi^2} \frac{\Omega_{\rm gw}(f)}{f^3}\,,
\ee
which differs from the one-sided strain spectral density 
$S_h(f)= (3 H_0^2/2\pi^2)\Omega_{\rm gw}(f)/f^3$ by a 
factor of $1/5$, since we are interested here in the strain 
response of a laser interferometer with a $90^\circ$ opening angle 
between the arms~\cite{RomanoCornish}.
Thus, we can rewrite the likelihood function \eqref{e:likefinal_colored_reduced} in
terms of $\Omega_{\rm gw}(f)$ and its optimal estimator $\hat\Omega_{\rm gw}(f)$ as 
\begin{multline}
p(d|\{\bar P_{1I}\}, \{\bar P_{2I}\},\Omega_{\rm gw})
\\
=\prod_\ell
\frac{e^{-2MN_{\rm seg}}}
{(\pi T/2)^{2MN_{\rm seg}}
%(T/4)^{2MN_{\rm seg}}
\prod_I(\bar P_{1I}(f_\ell) \bar P_{2I}(f_\ell))^M}
\exp\left[\frac{\hat \Omega_{\rm gw}^2(f_\ell)}
{2\bar\sigma_{\rm gw}^2(f_\ell)}\right]
\exp\left[-\frac{(\hat\Omega_{\rm gw}(f_\ell) 
-\Omega_{\rm gw}(f_\ell))^2}{2\bar\sigma^2_{\rm gw}(f_\ell)}\right]\,,
\label{e:likefinal_PE}
\end{multline}
where
\begin{align}
\frac{\hat\Omega_{\rm gw}(f_\ell)}{\bar\sigma^2_{\rm gw}(f_\ell)}
&\equiv\sum_I
\frac{2T\delta f}{\bar P_{1I}(f_\ell)\bar P_{2I}(f_\ell)}
\left(\frac{3H_0^2}{10\pi^2}\frac{\gamma(f_\ell)}{f_\ell^3}\right)
\frac{1}{(1+2/N_{\rm avg})}
%\nonumber\\
%&\hspace{1.25in}\times
\frac{2}{T}\frac{1}{M}\sum_{k=\ell-M/2}^{\ell+M/2-1}
{\rm Re}\left(\tilde d_{1I}^*(f_k) \tilde d_{2I}(f_k)\right)\,,
\label{e:Omega_gw_opt_freq}
\\
\frac{1}{\bar\sigma^2_{\rm gw}(f_\ell)}
&\equiv\sum_I
\frac{2T\delta f}{\bar P_{1I}(f_\ell)\bar P_{2I}(f_\ell)}
\left(\frac{3H_0^2}{10\pi^2}\frac{\gamma(f_\ell)}{f_\ell^3}\right)^2
\frac{1}{(1+2/N_{\rm avg})}\,.
\label{e:var_gw_opt_freq}
\end{align}
This is the form of the likelihood function that you'll find in
the LIGO-Virgo GWB literature, e.g., \cite{StochPE}, 
which serves as the starting point
for subsequent Bayesian parameter estimation analyses.

We can go one step further if we fix the spectral shape 
of the GWB, and focus attention on estimating only its 
amplitude at some reference frequency $f_{\rm ref}$, where
we normalize the spectral shape to have unit value.
(Typically $f_{\rm ref}=25~{\rm Hz}$ for LIGO-Virgo stochastic analyses.)
For example, for a power-law background with spectral index
$\alpha$, we have
\be
\Omega_{\rm gw}(f) \equiv \Omega_\alpha
\left(\frac{f}{f_{\rm ref}}\right)^\alpha\,,
\ee
(note that there is no implied sum over $\alpha$ in the above equation).
The spectral index of $\Omega_{\rm gw}(f)$ is related to the spectral index of $P_{h}(f)$ defined in the previous section by $\alpha=\beta+3$.
Then we can perform the product over frequencies $f_\ell$, again
using identity \eqref{e:xopt_identity} from Appendix~\ref{s:useful_identity}
to do the relevant sum of the argument of the exponential.
This yields a likelihood function and posterior distribution for
the amplitude $\Omega_\alpha$ that are both proportional to 
\be
\exp\left[-\frac{(\hat\Omega_\alpha -\Omega_\alpha)^2}{2\bar\sigma^2_\alpha}\right]\,,
\ee
where 
\begin{align}
\frac{\hat \Omega_\alpha}{\bar\sigma^2_\alpha}
&\equiv\sum_I
\frac{2T\delta f}{\bar P_{1I}(f_\ell)\bar P_{2I}(f_\ell)}
\left(\frac{3H_0^2}{10\pi^2}\frac{\gamma(f_\ell)}{f_{\rm ref}^3}
\left(\frac{f_\ell}{f_{\rm ref}}\right)^{\alpha-3}\right)
\frac{1}{(1+2/N_{\rm avg})}
%\nonumber\\
%&\hspace{1.25in}\times
\frac{2}{T}\frac{1}{M}\sum_{k=\ell-M/2}^{\ell+M/2-1}
{\rm Re}\left(\tilde d_{1I}^*(f_k) \tilde d_{2I}(f_k)\right)\,,
\label{e:Omega_alpha}
\\
\frac{1}{\bar\sigma^2_\alpha}
&\equiv\sum_I
\frac{2T\delta f}{\bar P_{1I}(f_\ell)\bar P_{2I}(f_\ell)}
\left(\frac{3H_0^2}{10\pi^2}\frac{\gamma(f_\ell)}{f_{\rm ref}^3}
\left(\frac{f_\ell}{f_{\rm ref}}\right)^{\alpha-3}\right)^2
\frac{1}{(1+2/N_{\rm avg})}\,.
\label{e:var_alpha}
\end{align}
Note that the factor 
\be
Q(f)\equiv
\frac{3H_0^2}{10\pi^2}\frac{\gamma(f_\ell)}{f_{\rm ref}^3}
\left(\frac{f_\ell}{f_{\rm ref}}\right)^{\alpha-3}
\frac{1}{\bar P_{1I}(f_\ell)\bar P_{2I}(f_\ell)}\,,
\ee
which multiplies the correlated data in \eqref{e:Omega_alpha}, 
is proportional to 
the standard expression for the {\it optimal filter} in the frequency 
domain, see e.g.,~\cite{Allen_Romano_1999, RomanoCornish}.
\end{widetext}

%%%%%%%%%%%%%%%%%%%%%%%%%%%%%%%%%%%%%%%%%%%
\subsection{Rigorous comparison}
\label{s:comparison}

As a check on the results of the previous section, we produce several
percentile-percentile (pp) plots \cite{pp_plot} to verify that the LIGO-Virgo
hybrid frequentist-Bayesian analysis has good statistical coverage.

To generate a pp plot, we first perform $N=300$ injections and recover the
injection with different likelihood functions, as described below. For each recovery, we
record the percentile of the posterior distribution at which the injected value
lies. We then plot the cumulative distribution function of the percentiles,
along with a 90\% credible interval determined using order statistics showing
the expected range of the cumulative distribution for each percentile value. If
the methods we use are unbiased, then the cumulative distribution should lie
within the 90\% credible interval, close to a diagonal line. Deviations from
this line indicate poor coverage, showing the method is biased.

\begin{figure}[htbp!]
\centering
\includegraphics[width=0.45\textwidth]{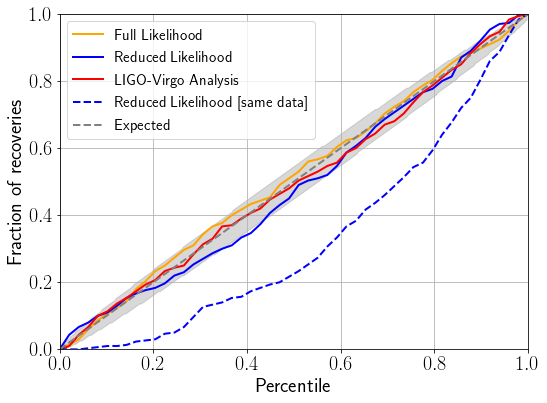} \caption{White signal+noise
analysis. The pp plot compares recoveries of the amplitude by the full likelihood, reduced 
likelihood and LIGO-Virgo stochastic analyses, showing that they both have good Bayesian
coverage. The latter two analyses are not identical because of different choices made
in conditioning the data. The dotted line shows the bias obtained when using 
the reduced likelihood analysis if the same data segment as the analysis segment
is used to estimate the auto-correlated power in the two detectors.} 
\label{fig:pp_plot_white}
\end{figure}

First, we consider a set of white signal+noise injections. 
We generate 300
noise and signal realizations, and recover the signal amplitude using the 
full Bayesian likelihood,
reduced likelihood for a white signal+noise model, 
and also the LIGO-Virgo hybrid frequentist-Bayesian analysis. 
As emphasized previously, for the reduced likelihood, it is important to use data segments different
from the analysis segment to estimate the auto-correlated power
in the two detectors, in order to avoid a bias in the recovered parameters. 
To illustrate
the importance of this point, we recover the signal using the reduced likelihood
analysis in two different ways: (i) using data segments different from the 
analysis segment to estimate the auto-correlated power, and 
(ii) using the same data segment as the analysis segment for 
estimating the auto-correlated power.
We show the results of those analyses in Figure~\ref{fig:pp_plot_white}. 
The analysis using the same data segment as the analysis segment 
to estimate the auto-correlated power is clearly biased. 
On the other hand, the full likelihood analysis, the reduced likelihood analysis and 
the LIGO-Virgo stochastic analysis, the latter two of which use data segments
different from the analysis segment to 
estimate the auto-correlated power, all show good coverage.
These analyses are not identical, however,
due to different conditioning of the data. In particular, the LIGO-Virgo
analysis computes the cross-correlation using 50\% overlapping Hann windows,
and estimates the auto-correlated 
power by averaging Welch estimators from adjacent data segments. 
In contrast, for our simple reduced likelihood analysis, 
we do not window the data (which is okay for white data), and use a single 
additional data segment for estimating the auto-correlated power.

Second, we perform a more realistic set of colored signal+noise injections, 
and recover the signal with the LIGO-Virgo stochastic analysis. In
these injections, the theoretical noise power spectra are kept the same,
although the noise realization changes in every injection. The amplitude of
$P_h(f)$ at a reference frequency of 1~Hz is fixed, but we draw the
spectral index from a uniform prior probability 
distribution ranging from $-3$ to $3$. We use the same prior on
the spectral index to perform the recovery. The realization of the GWB signal
changes in every injection. We perform two sets of 300 injections. In the
first set, the amplitude of $P_h(f)$ is set to be 1/8 of the amplitude of
$P_n(f)$ at the reference frequency in each segment, which is consistent with
the weak-signal approximation. In the second set, the amplitude of the signal is
set to 2 times the amplitude of the noise at the reference frequency in each
segment, which violates the weak-signal approximation.

\begin{figure}[htbp!]
\centering
\includegraphics[width=0.45\textwidth]{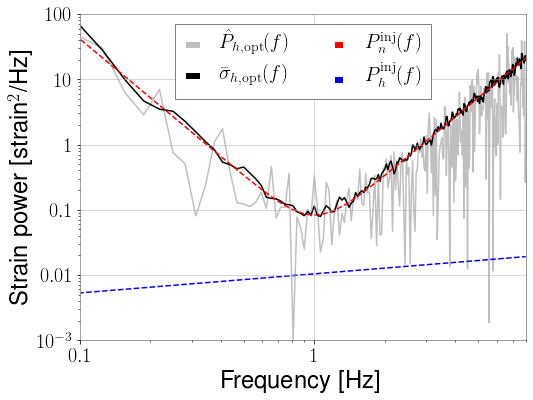}
\includegraphics[width=0.45\textwidth]{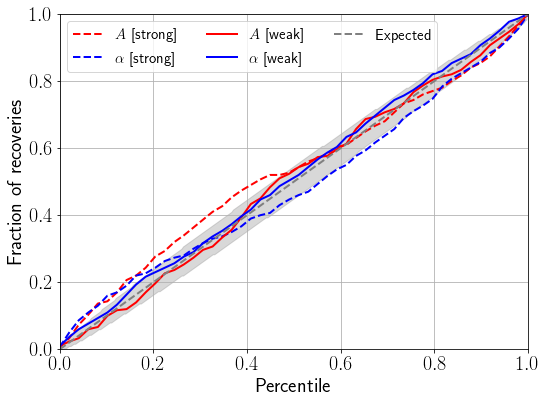}
\caption{Colored signal+noise analysis. In the top panel, we show the injected noise
power spectrum $P_n(f)$ (the same for both detectors)
and the injected GWB power spectrum $P_h(f)$, along with the optimal
estimator $\hat{P}_{h,{\rm opt}}(f)$ and its uncertainty $\bar{\sigma}_{h,\rm
opt}(f)$ for one segment in one realization. 
In the bottom panel, we show a pp
plot generated by performing 300 strong-signal and 
300 weak-signal injections and recoveries. 
We see that when the weak-signal approximation is satisfied,
the LIGO-Virgo stochastic analysis has excellent Bayesian coverage. Outside of
the weak-signal approximation, the coverage is less good, as expected.}
\label{fig:pp_plots_colored} 
\end{figure}

In the top panel of Figure~\ref{fig:pp_plots_colored}, we show the spectrum
associated with one injection in the weak-signal regime, along with the
injected noise and GWB signal spectra. In the bottom panel, we show the results
of the pp plot analysis for both strong and weak-signal amplitudes.
The recovery of both the amplitude and spectral index for weak signals 
lie within the 90\% uncertainty, demonstrating good Bayesian
coverage. On the other hand, the recovery of the amplitude and spectral index
for strong signals lies outside the 90\% region. This is expected,
given that strong signals violate the weak-signal approximation, which was 
used to derive the approximate equivalence of the LIGO-Virgo stochastic
analysis and the fully-Bayesian approach.

%% file: discussion.tex
\section{Discussion}
\label{s:discussion}

In this paper, we have shown in what sense LIGO-Virgo's 
hybrid frequentist-Bayesian cross-correlation analysis 
for a GWB is equivalent to a fully-Bayesian search.
The main result was our proof that for a reduced 
signal+noise model consisting of weak signals and 
estimated auto-correlated power spectra, the frequency 
integrand of the cross-correlation statistic and its
variance are sufficient statistics for the recovery of the GWB.
This means that the posterior distributions of the recovered 
spectrum of the GWB (e.g., its amplitude and spectral index)
will agree for the hybrid frequentist-Bayesian analysis 
and the fully-Bayesian analysis in the context of this 
reduced signal+noise model.
The results of our analyses on simulated data are 
consistent with those
found in \cite{Biscoveanu-et-al:2020}, which describes a 
fully-Bayesian implementation of LIGO's stochastic search 
that can estimate the presence of a primordial GWB in the
presence of an astrophysical foreground.

We note that a similar hybrid frequentist-Bayesian analysis 
is also being used for 
pulsar timing array searches for a GWB 
produced by inspiraling supermassive blackhole binaries 
associated with galaxy mergers.
Here a {\it noise-marginalized} version of the 
cross-correlation statistic~\cite{Vigeland-et-al:2018, Anholm-et-al:2009}
is used to avoid strong covariances 
that exist between individual estimates of pulsar red noise 
parameters and the amplitude of the GWB.
Noise marginalization is performed by drawing values of the 
pulsars' red noise parameters from posterior distributions 
that were generated by an earlier Bayesian analysis which 
jointly estimates the pulsar's red noise parameters and that 
of a common red process (e.g., the auto-correlated power due 
to the GWB), which may exceed the noise at the 
lowest observed frequencies.
This leads to a more accurate recovery of the amplitude of 
the GWB at relatively little computational cost to the 
cross-correlation statistic analysis.

A similar hierarchical approach, which first estimates the 
auto-correlated component of the background before looking
for evidence of cross-correlations, would mostly likely 
also be needed for analyzing
data from proposed {\it 3rd generation} (3G) ground-based 
detectors, like Cosmic Explorer~\cite{Reitze:2019_Cosmic_Explorer}
and Einstein Telescope~\cite{Maggiore:2020_ET}.
The expected sensitivity of these 3G detectors is such that
the weak-signal approximation in a single segment of data is
no longer valid, at least for the rate of binary black hole 
mergers that advanced LIGO and Virgo are currently 
observing~\cite{rates_O1_O2}.
As such, simply estimating the auto-correlated power spectra 
without assigning some portion of it to the GWB would lead to biased
estimates of the amplitude of the background, similar to 
what is seen in pulsar timing analyses~\cite{Vigeland-et-al:2018}.

It may be possible to extend the results of our paper to other 
signal+noise models, such as a 
non-stationary ``popcorn-like" background produced e.g., by 
stellar-mass binary black hole mergers in the LIGO frequency 
band~\cite{tbs_methods}.
For such a background, one would need to use a {\it mixture} 
signal+noise prior to model the non-stationarity, 
which amounts to postulating the presence of a correlated 
GWB signal in a certain fraction $\xi$ of the data segments and 
just noise for the remaining segments, with 
fraction $1-\xi$~\cite{tbs_methods}.
The detection of a GWB in this scenario would 
amount to a posterior distribution for $\xi$ strongly peaked 
away from zero.
By focusing on a GWB signal model that consists of just 
excess correlation in a 
certain fraction of data segments (as opposed to 
marginalizing over the parameters of 
potential BBH mergers~\cite{tbs_methods}),
one should be able to implement a computationally-efficient 
and robust (albeit suboptimal) search for a non-stationary GWB.

%% file: appA.tex
\section{Useful identity}
\label{s:useful_identity}

A useful identity that appears in several calculations in the 
paper is
\be
\sum_i \frac{(x_i-\mu)^2}{\sigma_i^2} 
= \sum_i \frac{x_i^2}{\sigma_i^2}
-\frac{x_{\rm opt}^2}{\sigma_{\rm opt}^2}
+\frac{(x_{\rm opt}-\mu)^2}{\sigma_{\rm opt}^2}\,,
\label{e:xopt_identity}
\ee
where $x_{\rm opt}$ and $\sigma_{\rm opt}^2$ are defined by
\be
\frac{x_{\rm opt}}{\sigma_{\rm opt}^2} 
\equiv \sum_i \frac{x_i}{\sigma^2_i}\,,
\qquad
\frac{1}{\sigma_{\rm opt}^2}\equiv 
\sum_i \frac{1}{\sigma^2_i}\,.
\ee
One can think of $x_i$ and $\sigma_i$, where $i=1,2,\cdots, N$, 
as a set of $N$ independent measurements and error bars of the
quantity $\mu$, whose value is to be determined from the measured data.
It is a well-known result that $x_{\rm opt}$ defined above is the 
{\it minimal-variance unbiased estimator} of $\mu$ with variance 
$\sigma_{\rm opt}^2$.
A proof of \eqref{e:xopt_identity} is the following:
\be
\begin{aligned}
\sum_i \frac{(x_i-\mu)^2}{\sigma_i^2} 
&= \sum_i \frac{x_i^2}{\sigma_i^2}
-2\mu\sum_i \frac{x_i}{\sigma_i^2}
+\mu^2\sum_i \frac{1}{\sigma_i^2}
\\
&= \sum_i \frac{x_i^2}{\sigma_i^2}
-2\mu\frac{x_{\rm opt}}{\sigma_{\rm opt}^2}
+\frac{\mu^2}{\sigma_{\rm opt}^2}
\\
&= \sum_i \frac{x_i^2}{\sigma_i^2}
-\frac{x_{\rm opt}^2}{\sigma_{\rm opt}^2}
+\frac{(x_{\rm opt}-\mu)^2}{\sigma_{\rm opt}^2}\,,
\end{aligned}
\ee
where we completed the square in $x_{\rm opt}$ and $\mu$ 
to get the last equality.

%% file: appB.tex
\section{Uncertainties in power spectrum estimates}
\label{s:psd_uncertainties}

In this appendix, we provide a brief summary of 
uncertainties in power spectrum estimates.
The final result of this analysis justifies the 
inclusion of the factor $(1+2/N_{\rm avg})$ 
in the expressions for the reduced likelihood
functions for both the white and colored
signal+noise models.
Our presentation follows that of an unpublished
internal LIGO technical note by Warren Anderson
(25 May 2004).

To simplify the notation a bit, we will use 
$\bar P_1$, $\bar P_2$ to denote
two power spectrum estimators, representing 
either the auto-correlations
$\bar\sigma_1^2$, $\bar\sigma_2^2$ for the 
white signal+noise model
or the auto-correlated power spectra 
$\bar P_{1I}(f_\ell)$, $\bar P_{2I}(f_\ell)$ for the 
colored signal+noise model. 
The number of averages used in the construction 
of the power spectrum estimators is denoted by
$N_{\rm avg}$, which is proportional
to the number of data samples $N$ for the white 
signal+noise models, or the number of 
frequency bins $M$ averaged over for coarse graining 
in a Welch power spectrum estimate~\cite{Welch:1967} 
for the colored signal+noise models.

Since $\bar P_1$ and $\bar P_2$ are unbiased 
estimators of $P_1$ and $P_2$, we can write
\be
\begin{aligned}
\bar P_1 = P_1 + \delta\bar P_1\,,
\\
\bar P_2 = P_2 + \delta\bar P_2\,,
\end{aligned}
\ee
where 
\be
\langle \delta\bar P_1\rangle=
\langle \delta\bar P_2\rangle= 0\,.
\ee
The variance and covariance of the power spectrum
estimators are given by the quadratic expectation 
values
\be
\begin{aligned}
&\langle(\delta\bar P_1)^2\rangle
= \langle \bar P_1^2\rangle - P_1^2 
\equiv {\rm var}(\bar P_1)\,,
\\
&\langle(\delta\bar P_2)^2\rangle
= \langle \bar P_2^2\rangle - P_2^2
\equiv {\rm var}(\bar P_2)\,,
\\
&\langle\delta\bar P_1\delta\bar P_2\rangle
= \langle \bar P_1\bar P_2\rangle - P_1 P_2
\equiv {\rm cov}(\bar P_1\bar P_2)\,.
\label{e:var_def1}
\end{aligned}
\ee
Explicitly evaluating $\langle \bar P_1^2\rangle$, etc., 
using the identity
\be
\langle a b c d\rangle
=\langle ab\rangle \langle cd\rangle
+\langle ac\rangle\langle bd\rangle
+\langle ad\rangle\langle bc\rangle
\ee
for zero-mean, Gaussian random variables, leads to
\be
\begin{aligned}
&\langle(\delta \bar P_1)^2\rangle = {P_1^2}/{N_{\rm avg}}\,,
\\
&\langle(\delta \bar P_2)^2\rangle = {P_2^2}/{N_{\rm avg}}\,,
\\
&\langle\bar P_1\bar P_2\rangle = {P_h^2}/{N_{\rm avg}}\,.
\label{e:var_def2}
\end{aligned}
\ee

Since the power spectra appear in the full likelihood 
function via the product of their inverses,  
$1/\sigma_1^2\sigma_2^2$ or $1/P_{1I}(f_\ell) P_{2I}(f_\ell)$, 
we need to calculate the expectation value
$\langle 1/\bar  P_1 \bar P_2\rangle$.
So making a Taylor series expansion
\be
\frac{1}{\bar P} = \frac{1}{P+\delta\bar P}
= \frac{1}{P}\left(1 - \frac{\delta\bar P}{P} 
+ \frac{\left(\delta\bar P\right)^2}{P^2} - \cdots\right)
\ee
for both $1/\bar P_1$ and $1/\bar P_2$, it follows that
\begin{widetext}
\be
\begin{aligned}
\frac{1}{\bar P_1\bar P_2} 
&= \frac{1}{P_1 P_2}
\left(1 - \frac{\delta\bar P_1}{P_1} + \frac{\left(\delta\bar P_1\right)^2}{P_1^2} - \cdots\right)
\left(1 - \frac{\delta\bar P_2}{P_2} + \frac{\left(\delta\bar P_2\right)^2}{P_2^2} - \cdots\right)
\\
&= \frac{1}{P_1 P_2}
\left(1 
- \frac{\delta\bar P_1}{P_1} 
- \frac{\delta\bar P_2}{P_2} 
+ \frac{\left(\delta\bar P_1\right)^2}{P_1^2} 
+ \frac{\left(\delta\bar P_2\right)^2}{P_2^2}
+ \frac{\delta\bar P_1\delta\bar P_2}{P_1 P_2} 
- \cdots\right)\,.
\end{aligned}
\ee
Taking the expectation value of both sides of the above expression, 
we find
\be
\bigg\langle\frac{1}{\bar P_1 \bar P_2}\bigg\rangle
= \frac{1}{P_1 P_2}\left(1 
+\frac{2}{N_{\rm avg}}
+\frac{1}{N_{\rm avg}}\frac{P_h^2}{P_1 P_2} - \cdots\right)
\simeq \frac{1}{P_1 P_2}\left(1 
+\frac{2}{N_{\rm avg}}\right)\,,
\ee
\end{widetext}
where we have ignored cubic and higher-order terms in
$\delta\bar P_1$ and $\delta\bar P_2$, and assumed the
weak-signal approximation, $P_h^2/P_1 P_2\ll 1$, to get the 
last approximately equality.
This result shows that $1/\bar P_1 \bar P_2$ is a 
{\it biased} estimator of $1/P_1 P_2$. 
Nonetheless, this bias can be removed by simply moving 
the factor of $(1+2/N_{\rm avg})$ to the left-hand side,
so that
\be
\frac{1}{\bar P_1 \bar P_2(1+2/N_{\rm avg})}
\ee
is a unbiased estimator of $1/P_1 P_2$.
This is the replacement we make for $1/P_1 P_2$ in the
reduced likelihood functions in the main text, cf.~\eqref{e:unbiased_1/P2}.

%% file: appC.tex
\section{Alternative derivation of a reduced likelihood function}
\label{s:alternative_derivation}

In this appendix, we give an alternative derivation of the 
likelihood function for a reduced signal+noise model, but
under different assumptions than those given in the 
main text.
More specifically, we do not assume here that the GWB signal 
is weak compared to the detector noise, nor do we use 
estimators of 
the auto-correlated power calculated from data segments 
different from that being analyzed for the signal.
Rather we assume that: (i) the number of data points $N$ 
(or coarse-grained averages $M$) for a given data segment $I$ 
is sufficiently large that one can expand the likelihood 
function around the maximum-likelihood estimators of 
$\sigma_1^2$, $\sigma_2^2$, $\sigma_h^2$ 
(or $P_{1I}(f_\ell)$, $P_{2I}(f_\ell)$, $P_h(f_\ell)$) 
to second order without loss of information; and
(ii) the data are informative for the auto-correlated power
$\sigma_1^2$, $\sigma_2^2$ (or $P_{1I}(f_\ell)$, $P_{2I}(f_\ell)$)
allowing us to evaluate the second-order likelihood function 
at the values of 
$\sigma_1^2$, $\sigma_2^2$ (or $P_{1I}(f_\ell)$, $P_{2I}(f_\ell)$)
that maximize the likelihood for {\it fixed} values of 
$\sigma_h^2$ (or $P_h(f_\ell)$).
For concreteness, we give the derivation here in the context 
of the white signal+noise model for two coincident and 
coaligned detectors. 
But it can be easily extended to the case of
colored data with non-stationary noise and 
a non-trivial overlap function.
Our derivation follows that given in 
\cite{Allen-et-al:2003_robust_bayesian}.

We start with the full 
likelihood function \eqref{e:likefinal_white} for the white
signal+noise model, which we rewrite here as
\be
p(d|\sigma^2_{n_1}, \sigma^2_{n_2}, \sigma_h^2)
=\exp\left[-\frac{N}{2} 
f(\sigma_1^2, \sigma_2^2, \sigma_h^2| d)\right]\,,
\label{e:p=exp(f)}
\ee
where
\begin{multline}
f(\sigma_1^2, \sigma_2^2, \sigma_h^2| d)=
2\ln(2\pi)+ \ln\left(\sigma_1^2\sigma_2^2-(\sigma_h^2)^2\right)
\\
\frac{1}{\left(\sigma_1^2 \sigma_2^2 -(\sigma^2_h)^2\right)}
\left(\hat\sigma_1^2 \sigma_2^2 + \hat\sigma_2^2\sigma_1^2
-2 \hat\sigma_h^2 \sigma_h^2\right)\,.
\end{multline}
In the above expressions,
$\sigma_1^2$, $\sigma_2^2$ 
are the total auto-correlated variances in the 
two detectors, see \eqref{e:S1S2_def}, and
\be
\begin{aligned}
&\hat \sigma^2_1 \equiv \frac{1}{N}\sum_i d_{1i}^2\,,
\\
&\hat \sigma^2_2 \equiv \frac{1}{N}\sum_i d_{2i}^2\,,
\\
&\hat \sigma^2_h \equiv \frac{1}{N}\sum_i d_{1i} d_{2i}\,,
\end{aligned}
\ee
are the maximum-likelihood estimators of 
$\sigma_1^2$, $\sigma_2^2$, $\sigma_h^2$.

Assuming that $N$ is sufficiently large, we can 
Taylor expand $f$ around its maximum-likelihood values
ignoring terms higher than second order in the 
differences 
$\sigma_1^2-\hat\sigma_1^2$, 
$\sigma_2^2-\hat\sigma_2^2$,
$\sigma_h^2-\hat\sigma_h^2$.
Doing so gives
\begin{widetext}
\be
f(\sigma_1^2, \sigma_2^2, \sigma_h^2| d)\simeq
f(\hat \sigma_1^2, \hat \sigma_2^2, \hat \sigma_h^2| d)
+ \frac{1}{2}\sum_{i,j} 
\frac{\partial^2 f}{\partial x_i\partial x_j}\bigg|_{\rm ML}
(x_i-\hat x_i)(x_j-\hat x_j)\,,
\label{e:f_approx}
\ee
where $x_i\equiv (\sigma_1^2,\sigma_2^2,\sigma_h^2)$,
and
\be
\frac{\partial^2 f}{\partial x_i\partial x_j}\bigg|_{\rm ML}
=\frac{1}{\left(\hat\sigma_1^2\hat\sigma_2^2-(\hat\sigma_h^2)^2\right)}
\left[
\begin{array}{ccc}
(\hat\sigma_2^2)^2 & (\hat\sigma_h^2)^2 & -2\hat \sigma_2^2\hat\sigma_h^2
\\
(\hat\sigma_h^2)^2 & (\hat\sigma_1^2)^2 & -2\hat \sigma_1^2\hat\sigma_h^2
\\
-2\hat\sigma_2^2\hat\sigma_h^2 & -2\hat\sigma_1^2\sigma_h^2 
& 2\left(\hat\sigma_1^2\hat\sigma_2^2+(\hat \sigma_h^2)^2\right)
\\
\end{array}
\right]\,.
\label{e:d2f/dx2}
\ee
Defining
\be
\Gamma_{ij} 
\equiv \frac{N}{2} 
\frac{\partial^2 f}{\partial x_i\partial x_j}\bigg|_{\rm ML}\,,
\label{e:fisher}
\ee
it follows that the inverse matrix
\be
C_{ij}\equiv \left(\Gamma^{-1}\right)_{ij}
=\frac{1}{N}
\left[
\begin{array}{ccc}
2(\hat\sigma_1^2)^2 & 2(\hat\sigma_h^2)^2 & 2\hat \sigma_1^2\hat\sigma_z^2
\\
2(\hat\sigma_h^2)^2 & 2(\hat\sigma_2^2)^2 & 2\hat \sigma_2^2\hat\sigma_z^2
\\
2\hat\sigma_1^2\hat\sigma_h^2 & 2\hat\sigma_2^2\sigma_h^2 
& \hat\sigma_1^2\hat\sigma_2^2+(\hat \sigma_h^2)^2
\\
\end{array}
\right]
\label{e:cov_fisher}
\ee
is an estimator of the covariance matrix of the maximum-likelihood
estimators $\hat\sigma_1^2$, $\hat\sigma_2^2$, $\hat\sigma_h^2$.

To proceed further, we construct a {\it reduced} likelihood 
function by assuming that the data are informative with respect
to the detector auto-correlations $\sigma_1^2$, $\sigma_2^2$.
This means that we can evaluate \eqref{e:f_approx} at the 
values of $\sigma_1^2$, $\sigma_2^2$ 
that maximize \eqref{e:f_approx} for {\it fixed} values of 
$\sigma_h^2$.
So, simultaneously solving the two equations
\be
\frac{\partial f}{\partial\sigma_1^2} = 0\,,
\qquad
\frac{\partial f}{\partial\sigma_2^2} = 0\,,
\ee
for $\sigma_1^2$, $\sigma_2^2$, we obtain
\be
\begin{aligned}
\sigma_1^2 = \hat\sigma_1^2 + 
\frac{2\hat\sigma_1^2\hat\sigma_h^2}
{\left(\hat\sigma_1^2\hat\sigma_2^2 + (\hat\sigma_h^2)^2\right)}
(\sigma_h^2-\hat\sigma_h^2)\,,
\qquad
\sigma_2^2 = \hat\sigma_2^2 + 
\frac{2\hat\sigma_2^2\hat\sigma_h^2}
{\left(\hat\sigma_1^2\hat\sigma_2^2 + (\hat\sigma_h^2)^2\right)}
(\sigma_h^2-\hat\sigma_h^2)\,.
\label{e:autocorrelations_reduced}
\end{aligned}
\ee
Denoting the RHSs of these expressions by 
$\bar\sigma_1^2$, $\bar\sigma_2^2$, it follows that \eqref{e:f_approx} becomes
\be
f(\sigma_1^2, \sigma_2^2, \sigma_h^2|d)\Big|_{\sigma_1^2 = \bar\sigma_1^2,\,
\sigma_2^2=\bar\sigma_2^2}\simeq
f(\hat\sigma_1^2, \hat\sigma_2^2, \hat\sigma_h^2|d)
+\frac{(\sigma_h^2-\hat\sigma_h^2)^2}
{\left(\hat\sigma_1^2\hat\sigma_2^2 + (\hat\sigma_h^2)^2\right)}\,.
\ee
The corresponding reduced likelihood function is
\be
p(d|\bar\sigma_1^2,\bar\sigma_2^2,\sigma_h^2) \equiv
p(d|\hat\sigma_1^2,\hat\sigma_2^2, \hat\sigma_h^2)\,
\exp\left[-\frac{N}{2}
\frac{(\sigma_h^2-\hat\sigma_h^2)^2}
{\left(\hat\sigma_1^2\hat\sigma_2^2 + (\hat\sigma_h^2)^2\right)}
\right]\,.
\label{e:reduced_like_alt}
\ee
Thus, we see that $\hat\sigma_h^2$ together with $\hat\sigma_1^2$, 
$\hat\sigma_2^2$ are sufficient statistics for $\sigma_h^2$ with variance 
\be
{\rm var}[\hat\sigma_h^2] = \left(\hat\sigma_1^2\hat\sigma_2^2 + (\hat\sigma_h^2)^2\right)/N\,.
\label{e:reduced_var_alt}
\ee
\end{widetext}

A couple of remarks are in order:

(i) Nowhere in the above derivation did we assume that the power 
in the GWB is small compared to the detector noise.
Thus, the reduced likelihood function \eqref{e:reduced_like_alt} 
is valid for arbitrarily large GWB signals, which is relevant, for 
example, for searches for a GWB using pulsar timing 
arrays~\cite{Siemens-et-al:2013, arzoumanian2020nanograv}
or the proposed space-based interferometer 
LISA~\cite{baker2019laser}; see e.g.,~\cite{Adams:2013qma}.
Equation \eqref{e:reduced_var_alt} 
contains an extra term, $(\hat\sigma_h^2)^2$, compared to
\eqref{e:reduced_var_white}, which takes into 
account the extra variance associated with the GWB, over and
above that which is already captured in the auto-correlations
$\hat\sigma_1^2$ and $\hat\sigma_2^2$.

(ii) Although we denoted the right-hand sides of 
\eqref{e:autocorrelations_reduced}
by $\bar\sigma_1^2$, $\bar\sigma_2^2$, these expressions are not
the same as those used in the main text (see the discussion in
Sec.~\ref{s:white_reduced}), which were estimators of 
$\sigma_1^2$, $\sigma_2^2$ constructed from data segments 
{\it different} than that being analyzed for the signal.
The use of different data segments to estimate the auto-correlations
is necessary for LIGO-Virgo data, since the number of averages used 
to estimate power spectra is not sufficiently large to beat down the
bias that arises from using the same data in both the numerator and 
denominator of the exponential in \eqref{e:reduced_like_alt}.
(Recall that for the colored signal+noise model, the number of 
averages is proportional to the number of
frequency bins $M$ that are averaged together for coarse-graining.)  
For LIGO-Virgo data, one is restricted to $N_{\rm avg}$ of order at 
most 50, due to the complexity of the detector noise (the need to take
the coarse-grained frequency resolution to be 
$\delta f\sim 0.25~{\rm Hz}$) and its broad non-stationarity on
time scales of order minutes (ignoring shorter time-scale 
instrumental glitches).
By using different data segments to estimate the auto-correlated
power, the bias goes away in the weak-signal limit, as shown 
in Fig.~\ref{fig:pp_plot_white}.
But if the number of averages was sufficiently large, then one could
use the expression in \eqref{e:reduced_like_alt} as is.